\documentclass[aps,prd,twocolumn,nofootinbib,superscriptaddress]{revtex4-1}

\usepackage{amssymb}
\usepackage{amsmath}
\usepackage{multirow}
\usepackage{mathrsfs}
\usepackage{graphicx}
\usepackage{color}
\usepackage{dcolumn}
\usepackage{bm}
\newfont{\Fr}{eufm10}
\usepackage{graphics}
\usepackage{epsfig}
\usepackage[english]{babel}
\usepackage{verbatim}
\usepackage{amsfonts}
\usepackage[latin1]{inputenc}
\usepackage{hyperref}
\usepackage{soul}

\RequirePackage[english]{babel}
\RequirePackage{amsmath} % Useful for equation alignment
\RequirePackage{verbatim} % Allows to have comments over multiple lines
\RequirePackage{mathrsfs}
\RequirePackage{amsfonts}
\RequirePackage{amssymb}
\RequirePackage[latin1]{inputenc}
\RequirePackage{hyperref}

\newcommand{\tc}{\eta_k^c}
\newcommand{\beq}{\begin{equation}}
\newcommand{\eeq}{\end{equation}}
\newcommand{\barr}{\begin{eqnarray}}
\newcommand{\earr}{\end{eqnarray}}
\newcommand{\bea}{\begin{eqnarray*}}
\newcommand{\eea}{\end{eqnarray*}}
\newcommand{\nk}{\textbf{k}}
\newcommand{\dphi}{\delta \phi}

\newcommand{\x}{\textbf{x}}

\newcommand{\bra}{\langle}
\newcommand{\ket}{\rangle}
\newcommand{\mH}{\mathcal{H}}

\newcommand{\nn}{\nonumber \\}

\newcommand{\half}{\frac{1}{2}}

\newcommand{\cre}{\hat{a}^{\dagger}}    % creation

\newcommand{\ann}{\hat{a}}

\newcommand{\mR}{\mathcal{R}}

\begin{document}

\title{Novel vacuum conditions in inflationary collapse models}
\author{Gabriel R. Bengochea}
\email{gabriel@iafe.uba.ar} \affiliation{Instituto de Astronom\'\i
a y F\'\i sica del Espacio (IAFE), CONICET - Universidad de Buenos Aires, (1428) Buenos Aires, Argentina}
\author{Gabriel Le\'{o}n}
\email{gleon@fcaglp.unlp.edu.ar} \affiliation{Grupo de Astrof\'\i sica, Relatividad y Cosmolog\'\i a,
Facultad de Ciencias Astron\'{o}micas y Geof\'\i sicas, Universidad Nacional de La Plata,
Paseo del Bosque S/N (1900) La Plata, Argentina}

\begin{abstract}

Within the framework of inflationary models that incorporate a spontaneous reduction of the wave function for the emergence of the seeds of cosmic structure, we study the effects on the primordial scalar power spectrum by choosing a novel initial quantum state that characterizes the perturbations of the inflaton. Specifically, we investigate under which conditions one can recover an essentially scale free spectrum of primordial inhomogeneities when the standard Bunch-Davies vacuum is replaced by another one that minimizes the renormalized stress-energy tensor via a Hadamard procedure. We think that this new prescription for selecting the vacuum state is better suited for the self-induced collapse proposal than the traditional one in the semiclassical gravity picture. We show that the parametrization for the time of collapse, considered in previous works, is maintained. Also, we obtain an angular spectrum for the CMB temperature anisotropies consistent with the one that best fits the observational data. Therefore, we conclude that the collapse mechanism might be of a more fundamental character than previously suspected.

\end{abstract}

\pacs{Valid PACS appear here}


\maketitle

\section{Introduction}\label{intro}

Inflation is considered as a fundamental component of the standard $\Lambda$CDM cosmological model characterizing the initial stages of the universe \cite{starobinsky,guth,linde,albrecht}. Essentially, according to the inflationary paradigm, the early universe underwent an accelerated expansion induced by a scalar field named the inflaton. In addition, it is widely accepted that the quantum fluctuations of the inflaton gave birth to the primordial curvature perturbation, which in turn, generated the primeval density perturbations \cite{mukhanov81,mukhanov2,starobinsky2,hawking,hawking2}. These primordial perturbations are thus responsible for the origin of all the observed structure in the universe. The predicted properties of such perturbations are consistent with recent observational data from the cosmic microwave background (CMB) \cite{planck2015,planck2015likelihoods,plkinflation15}. In particular,
the data are consistent with a nearly scale invariant spectrum associated to the perturbations, which also favors the simplest inflationary models \cite{plkinflation15,Martin14}.

According to the standard inflationary picture, the dynamical expansion of the early universe is governed by Einstein equations which are symmetry preserving; the symmetry being the homogeneity and isotropy. Another important aspect is that, when considering the quantum description of the fields, the vacuum state associated to the inflaton is also homogeneous and isotropic, i.e. it is an eigenstate of the operator generating spatial translations and rotations. Furthermore, the dynamical evolution of the vacuum satisfies the Schr\"odinger equation, which does not break translational and rotational invariance. As a consequence, we arrive at an important conundrum: it is not clear how from a perfect symmetric initial situation (both in the background spacetime and in the quantum state that characterizes the inflaton), and based on dynamics that preserves the symmetries (the homogeneity and isotropy), one ends up with a final state that is inhomogeneous and anisotropic describing the late observed universe. The aforementioned problem was originally introduced in \cite{PSS06} (and extensively discussed in \cite{Shortcomings,LLS13})  together with  a possible solution: the self-induced collapse hypothesis. The collapse proposal consists that at some point, during the inflationary epoch, a spontaneous change occurs, transforming the original quantum state of the inflaton (the vacuum) into a new quantum state lacking the symmetries of the initial state.

It is worthwhile to mention that the situation we are facing is connected with the so called quantum measurement problem. Sometimes in the literature, the problem is presented as the quantum-to-classical transition of the primordial quantum fluctuations, and then decoherence is introduced into the picture \cite{polarski,kiefer}. Although, decoherence can provide a partial understanding of the issue, it does not fully addresses the problem mainly because decoherence does not solve the quantum measurement problem. We will not dwell in all the conceptual aspects regarding the appeal of decoherence during inflation, neither the perceived advantages that the objective reduction models could offer when applied to the early universe. Instead, we referred the interested reader to Refs. \cite{PSS06,Shortcomings} for a more in depth analysis.

The collapse hypothesis during inflation has been analyzed using two approaches: In the first, one characterizes the post-collapse state phenomenologically through the expectation values and quantum uncertainties of the field, and its conjugated momentum, evaluated at the time of collapse \cite{PSS06,US08,DT11,pia}. In the second approach, one employs a particular collapse mechanism called the continuous spontaneous localization model, where a modification of the Schr\"odinger equation is proposed, resulting in an objective dynamical reduction of the wave function \cite{CPS13,jmartin,tpsingh,LB2015,magueijo2016}. In both approaches,  one obtains a prediction for the scalar and tensor power spectra that, in principle, is different from the standard prediction \cite{lucila,abhishek}. The first approach has been tested using the most recent data provided by the Planck collaboration, and, under certain circumstances, provides the same  Bayesian evidence of the minimal standard cosmological model $\Lambda$CDM \cite{benetti06}. Therefore, we will follow the first approach to characterize the self-induced collapse, but the framework exposed in the present work can be extended to the second approach.

Another important feature of the collapse proposal is the adoption of semiclassical gravity  \cite{WaldQFTCS}, which serves to relate the spacetime description in terms of the metric and the degrees of freedom of the inflaton. In the semiclassical picture, gravity is always classical while the matter fields are treated quantum mechanically. We assume such a framework to be a valid approximation during the inflationary era, which is well after the full quantum gravity regime has ended. This approach is different from the standards models of inflation in which metric and matter fields are quantized simultaneously. We should mention that there are many arguments suggesting that the spacetime geometry might emerge from deeper, non-geometrical and fundamentally quantum mechanical degrees of freedom \cite{Em1,Em2,Em3,Em4,Em5}. Therefore, in this work, we will employ Einstein semiclassical equations $G_{ab} = 8 \pi G \bra \hat{T}_{ab} \ket$.

On the other hand, the selection of the pre-collapse state, i.e. the vacuum state, which is perfectly homogeneous and isotropic, is not generic. It is known that since we are dealing with a theory of a scalar field (the inflaton) in a curved spacetime, the choice of the vacuum state is not unique \cite{WaldQFTCS,fullingbook}. Traditionally, the Bunch-Davies (BD) vacuum is selected when considering the quantum theory of the inflaton. The criterion used for the BD vacuum is based on finding a state $|0\ket$ such that it minimizes the expectation value $\bra  0 | \hat H (\eta_i) | 0 \ket$ at some initial time $\eta_i$, with $\hat H$ the Hamiltonian associated to the perturbations \cite{Mukhanov1992,MukhanovQFTCS}; this prescription is also called Hamiltonian diagonalization. On the other hand, there are known unresolved issues with such procedure. One is that $\bra  0 | \hat H (\eta_i) | 0 \ket$ can be minimized only at an instant $\eta_i$; at some other time $\eta_1 > \eta_i$, the BD vacuum does not achieve the sought minimization of the expectation value. In other words, the zero ``particle'' state is only defined at the time $\eta_i$, and as inflation unfolds, the state $| 0 \ket$ contains ``particles'' at other time $\eta_1$. Another related issue is that usual renormalization methods, which make $\bra  0 | \hat H (\eta_i) | 0 \ket$  finite, can only be defined at $\eta_i \to -\infty$, that is, at the very early stages of inflation. Some authors consider those arguments sufficient to find alternatives to the Hamiltonian diagonalization method \cite{fulling79,picon2007,handley16}. Here it is also important to mention that different choices other than the BD vacuum state have been analyzed previously. For example in Refs.  \cite{Martin2000,Danielsson2002,Martin2003} it is presented an analysis regarding the observable effects of trans-Planckian physics in the CMB and its relation with a non-BD vacuum. In addition, a non-BD vacuum is usually associated with large non-Gaussianities in the CMB \cite{Komatsu2010,Agullo2011}.

One of the possible alternatives is proposed in Ref. \cite{handley16}. Those authors suggest that, instead of minimizing $\bra  0 | \hat H | 0 \ket$, one should focus on minimizing the renormalized $\bra  \hat T_{00} (x) \ket$. Specifically, the vacuum $|\tilde 0 \ket$ (which is not the same as the BD vacuum), is such that it minimizes the 0-0 component of the renormalized expectation value of the energy-momentum tensor, which can be considered as a local energy density. Moreover, the
vacuum $|\tilde 0 \ket$ only minimizes the renormalized $\bra \tilde 0 | \hat T_{00} (x) | \tilde 0 \ket$ at some particular time $\eta_0$. However, conceptually, it is easier to handle a notion of an instantaneous local energy density minimum than dealing with a notion of ``particle'' that changes with time. Also, the time $\eta_0$ does not need to be taken in the limit $\eta_0 \to -\infty$, although, if one chooses to set $\eta_0$ at such early times, then $|\tilde 0 \ket$ coincides with the prescription of the BD vacuum, but not with its physical interpretation of a ``particle-less'' state.

All previous works regarding the self-induced collapse proposal, when applied to the inflationary scenario,  have been based on selecting the BD vacuum, which is the usual choice in traditional models of inflation as well. Nonetheless, one of the key objects in the inflationary collapse proposal, based on the semiclassical gravity framework,  is the expectation value $\bra  \hat T_{ab} (x) \ket$. In our approach, if the post-collapse state does not share the same symmetries as the initial--vacuum--state then $\bra \hat{T}_{ab} \ket$, evaluated in the post-collapse state, will result in a geometry that is no longer homogeneous and isotropic, thus, providing the primordial perturbations for cosmic structure. Therefore, a criterion based on selecting a vacuum state that minimizes the renormalized expectation value of $ \hat T_{00}$ seems better suited for our picture than one based on choosing a zero ``particle'' state at a particular time. Furthermore, after the collapse, clearly  $\bra \hat T_{00} \ket$ will no longer be the same   as the one evaluated in the vacuum state. Hence, if one thinks the collapse as a dynamical process, changing continuously from $|\tilde 0 \ket$ to the post-collapse state, then it is clear to picture the expectation value of $\hat T_{00}$ also  changing continuously. In particular, the value  $\bra \tilde 0 | \hat T_{00}| \tilde 0 \ket$ will transform from a minimum, which produces a perfectly symmetric spacetime, into a different value generating the perturbations of the geometry.

From discussion above the motivation for the present work is established. That is, we are interested in analyzing the possible effects on the primordial power spectrum generated by choosing the novel prescription based on minimizing the renormalized $\bra \hat T_{00} \ket$. In particular, we are interested in analyzing which aspects of the collapse proposal are modified when the initial conditions are also changed. As we will show, one of our findings indicate that the parametrization of the time of collapse,  for each mode of the field, surprisingly remains the same. This led us to think that the physics behind the self-induced collapse of the wave function should be studied in more detail.

The article is organized as follows: in Sect. \ref{secdos}, we review some basics about inflation in the semiclassical gravity framework; in Sect. \ref{sectres}, we analyze the quantization of perturbations, the vacuum choice and present the emergence of curvature perturbation within the collapse hypothesis. Then, we show our prediction for the scalar power spectrum. In Sect. \ref{seccuatro} we make a discussion of our results, and finally in Sect. \ref{conclusions} we summarize our conclusions.

Regarding conventions and notation, we will be using a $(-,+,+,+)$ signature for the spacetime metric, and we will use units where $c=1=\hbar$.

\section{Inflation in the semiclassical picture}
\label{secdos}

In this section, we will summarize some basic concepts regarding the inflationary
model in the framework of semiclassical gravity. Extra details can be consulted in previous works (e.g. \cite{PSS06, US08,Leon10,DLS11}).

In the inflationary regime, the dominant type of matter is modeled by a scalar field $\phi$, called the inflaton, with a potential $V$ responsible for the
accelerating expansion. At the end of the inflationary epoch, the universe follows the standard Big Bang evolution whose transition mechanism
is provided by a reheating period.

We begin describing the inflationary universe by the action of a scalar field minimally coupled to gravity,
\beq\label{actioncol}
S[\phi,g_{ab}]=\int d^4x \sqrt{-g} \bigg[ \frac{1}{16 \pi G} R[g] - \half \nabla_a \phi \nabla_b \phi g^{ab} - V[\phi] \bigg].
\eeq
Varying this last equation with respect to the metric yields the Einstein field
equations $G_{ab} = 8 \pi G T_{ab}$, with $G_{ab}$ the Einstein tensor.

We will use conformal coordinates and, as usual,  we will split the metric and the scalar field into a background perfectly homogeneous and isotropic, plus small perturbations. That is, we write the metric as $g_{ab} = g_{ab}^{(0)} + \delta g_{ab}$, and $\phi = \phi_0 (\eta) + \dphi (\x,\eta)$, where the background will be represented by a spatially flat FLRW spacetime and the homogeneous part of the scalar field (in the slow-roll regime) by $ \phi_0 (\eta) $. From Einstein equations for the background, it follows that $G_{00}^{(0)}=8\pi G T_{00}^{(0)}=8\pi G a^2 \rho$, so the Friedmann equation is $3\mH^2 = 8 \pi G a^2 \rho$ where $\mH \equiv a'(\eta)/a(\eta)$ is the conformal Hubble parameter, and $a(\eta)$ is the scale factor. As is customary, the scale factor will be set to $a=1$ at the present time. Remember that the inflationary phase extends between $-\infty < \eta < \eta_r$, where $\eta_r\approx -10^{-22}$ Mpc is the conformal time when inflation comes to an end. From here on, primes over functions will denote derivatives with respect to the conformal time $\eta$. During the inflationary phase, the potential $V$ is the major contribution to the energy density $\rho$.

In the slow-roll inflationary model, the conformal Hubble parameter is expressed by $\mH \simeq {-1}/{[\eta(1-\epsilon_1)]}$, with $ \epsilon_1 \equiv 1 -
{\mH'}/{\mH^2}$ the Hubble slow-roll parameter, which during inflation $1\gg \epsilon_1 \simeq$ constant.

We will only focus on first-order scalar perturbations; hence, the FLRW perturbed metric can be written as
\barr
ds^2 &=& a^2 (\eta) \big\{  -(1-2\varphi) d\eta^2 + 2 (\partial_i B) dx^i d\eta
+\nonumber \\
&+& [ (1-2\psi) \delta_{ij} + 2 \partial_i \partial_j E] dx^i dx^j \big\}.
\earr
Within the semiclassical framework, it is convenient to work with the well known gauge-invariant Bardeen potentials. They are defined as $\Phi \equiv \varphi + \frac{1}{a} [a (B-E')]'$ and $\Psi \equiv \psi +\mH (E'-B) $. On the other hand, the inflaton perturbation can be modeled by the gauge-invariant fluctuation of the scalar field $\dphi^{(\textrm{GI})}(\eta,\x) = \dphi + \phi_0' (B-E')$.

Working with the perturbed Einstein equations (in the absence of anisotropic stress), it can be found that $\Psi = \Phi$. Also, these perturbed equations, along with the Friedmann equation and the equation of motion for $\phi_0$ in the slow-roll approximation, i.e. $3\mH \phi_0' + a^2 \partial_\phi V \approx 0$, imply that (see for instance Appendix A of \cite{gabrielqdesitter}):
\beq\label{25x}
\nabla^2 \Psi +\mu \Psi =  4 \pi G \phi_0' \dphi^{'(\textrm{GI})},
\eeq
where $\mu \equiv \mH^2-\mH'=\epsilon_1 \mH^2$. In Fourier space, Eq. \eqref{25x} results,
\beq\label{25b2}
\Psi_{\nk} (\eta) = \sqrt{\frac{\epsilon_1}{2}} \frac{H}{M_P (k^2-\mu)} a
\dphi'_{\nk} (\eta)^{(\textrm{GI})},
\eeq
with $H$ the Hubble parameter, $M_P\equiv \sqrt{1/(8\pi G)}$ is the reduced Planck mass, and we have also used the definition of $\epsilon_1$. Notice that during most of the inflationary phase, the inequality $k^2 \gg \mu$ is satisfied (both when $|k\eta| \gg 1$ and $|k \eta| \ll 1$). Only when $\epsilon_1$ starts departing from being a constant (i.e. when $\epsilon_1\to 1$ which means that inflation is ending) that inequality is violated. Given that modes of observational interest are bigger than the Hubble radius ($|k\eta| \ll 1$) while the inflationary phase is still going on, the approximation $k^2 \gg \mu$ remains valid. Therefore, Eq. \eqref{25b2} can be approximated by
\beq\label{master0}
\Psi_{\nk} (\eta) \simeq  \sqrt{\frac{\epsilon_1}{2}} \frac{H}{M_P k^2} a
\dphi'_{\nk} (\eta)^{(\textrm{GI})}.
\eeq

In the semiclassical framework, Eq. \eqref{master0} can be generalized to
\beq\label{master}
\Psi_{\nk} (\eta) \simeq  \sqrt{\frac{\epsilon_1}{2}} \frac{H}{M_P k^2} a \bra
\hat{\dphi'}_{\nk} (\eta)^{(\textrm{GI})} \ket.
\eeq
Last equation is expressed in terms of gauge-invariant quantities $\Psi_{\nk} (\eta)$ and $\hat{\dphi'}_{\nk} (\eta)^{(\textrm{GI})}$. Note that, in our approach, the metric perturbation will be always a classical quantity.

\section{Quantum perturbations, vacuum choice and collapse hypothesis}
\label{sectres}

In this section, we perform the quantization of the perturbations. However, before proceeding with the quantization, we will first briefly address the subject of gauge and its relation with the metric and field perturbations.

In the following, we will choose to work with a fixed gauge and not in terms of the so-called gauge-invariant combinations. We are forced to do so because, in our approach, the adoption of the semiclassical gravity framework leads to consider a classical metric perturbation and a quantum field perturbation, i.e. the metric and field perturbations are treated on a different footing. This contrasts with the standard treatment in which, normally, one chooses to work with gauge invariant quantities which mix matter and geometry degrees of freedom. Then, the quantization results essentially the same for both types of perturbations (matter and geometry).

On the other hand, the choice of gauge implies that the time coordinate is attached to some specific slicing of the perturbed spacetime. And thus, our identification of the corresponding hypersurfaces, those of constant time as the ones associated with the occurrence of collapses--something deemed as an actual physical change--turns what is normally a simple choice of gauge into a choice of the distinguished hypersurfaces, tied to the putative physical process behind the collapse. This naturally leads to tensions with the expected general covariance of a fundamental theory, a problem that afflicts all known collapse models, and which in the non-gravitational settings becomes the issue of compatibility with Lorentz or Poincare invariance of the proposals. We must acknowledge that this generic problem of collapse models is indeed an open issue for the present approach. One would expect that its resolution would be tied to the uncovering of the actual physics behind what we treat here as the collapse of the wave function (which we  view as a merely effective description). As it has been argued in related works, and in ideas originally exposed by Penrose \cite{penrose1996}, we hold that the physics that lies behind all this links the quantum treatment of gravitation with the foundational issues afflicting quantum theory in general; and in particular, those with connection to the so-called ``measurement problem''.

The gauge we choose is the \emph{longitudinal gauge} ($B=E=0$). The advantage of working with this gauge is that the action at second order involving the matter and metric perturbations is mathematically the same as the one using gauge invariant quantities, i.e. the Bardeen potentials and $\dphi^{(\textrm{GI})}$. Also, note that $\Psi$ represents the curvature perturbation in this gauge, and it is related to $\dphi$ in the exact same way as in Eq. \eqref{master} \cite{brandenberger1993}. Therefore, we can be certain that the field perturbations are actual physical degrees of freedom and not pure gauge. Additionally, in our approach, before the collapse (i.e. in the vacuum state) there are no metric perturbations. Hence, the resulting action is the one involving only $\dphi$. After the collapse, when the metric perturbations are indeed present, the quantum theory should be modified as presented in \cite{DT11}. However, we will not consider such backreaction, mainly because we are interested in describing the quantum theory using a non-BD vacuum.

Next, we will present the quantum theory for the field $\dphi (\x,\eta)$, which will be carried out by choosing a vacuum state different from the BD vacuum. We point out that the criterion used is physically different from the usual BD vacuum. Then, we will characterize the collapse scheme, calculate the curvature perturbation, and finally we will show our expression for the primordial scalar power spectrum.

We start (for simplicity) re-scaling the field variable as $y=a\dphi$. Then, we proceed by expanding the action \eqref{actioncol} up to second order in the scalar field perturbation $y$. This results in:
\barr\label{acciony}
\delta S^{(2)} &=&  \int d^4x \frac{1}{2}
\bigg[ y'^2 - (\nabla y)^2 + \left(\frac{a'}{a} \right)^2 y^2 \nonumber  \\
&-& 2 \left(\frac{a'}{a} \right) y y' -y^2a^2  \partial_{\phi \phi}^2 V \bigg].
\earr
Therefore, the canonical momentum conjugated to $y$ is $\pi \equiv \partial \delta
\mathcal{L}^{(2)}/\partial y' = y'-(a'/a)y=a\dphi'$.

In order to facilitate the calculations, we will neglect the slow roll parameters $\epsilon_1$ and $\epsilon_2\equiv \epsilon_1'/(\mH \epsilon_1)$ in the quantization procedure. At the end of the computations, we will argue how we can generalize our result to the quasi-de Sitter case, in which the slow roll parameters are considered.

Now, the field and momentum variables are promoted to operators satisfying the equal time commutator relations
$[\hat{y}(\x,\eta), \hat{\pi}(\x',\eta)] = i\delta (\x-\x')$ and $[\hat{y}(\x,\eta), \hat{y}(\x',\eta)] = 0= [\hat{\pi}(\x,\eta), \hat{\pi}(\x',\eta)] $.
Expanding the fields operators in Fourier modes yields
\barr
\hat{y}(\eta,\x) &=& \frac{1}{L^3} \sum_{\nk} \hat{y}_{\nk} (\eta) e^{i \nk \cdot\x} \\
\hat{\pi}(\eta,\x) &=& \frac{1}{L^3} \sum_{\nk} \hat{\pi}_{\nk} (\eta) e^{i \nk \cdot\x}
\earr
where the sums are over the wave vectors $\nk$, satisfying $k_i L=2\pi n_i$ for $i=1,2,3$ with $n_i$ integer. Also, we have defined $\hat y_{\nk} (\eta) \equiv y_k(\eta)
\ann_{\nk} + y_k^*(\eta) \cre_{-\nk}$ and  $\hat \pi_{\nk} (\eta) \equiv
g_k(\eta) \ann_{\nk} + g_{k}^*(\eta) \cre_{-\nk}$, with $g_k(\eta) = y_k'(\eta)
- \mH y_k (\eta)$ and $\ann_{\nk}, \cre_{\nk}$ being the usual annihilation/creator operators, respectively. Note that the quantization is on a finite cubic box of length $L$, and at the end of the calculations we will take the continuum limit ($L \to \infty$, $\nk \to $ cont.).

From action \eqref{acciony}, the equation of motion for  $y_k(\eta)$ results in
\beq\label{ykmov2}
y''_k(\eta) + \left(k^2 - \frac{a''}{a} \right) y_k(\eta)=0
\eeq
with $a''/a = 2/\eta^2 $. The general solution is,
\beq\label{ykgensol}
y_k(\eta)= A_k \Big(1-\frac{1}{k\eta}\Big) e^{-i k\eta}+B_k\Big(1+\frac{1}{k\eta}\Big) e^{i k\eta}
\eeq
where $A_k$ and $B_k$ are two constants (dependent on $k$) that will be fixed by the initial conditions at some $\eta_0$.

Therefore, to complete the quantization, we have to specify the solutions $y_k(\eta)$, through constants $A_k$ and $B_k$. This choice is not completely free; to insure that
canonical commutation relations between $\hat y$ and $\hat \pi$ give $[\hat{a}_{\nk},\hat{a}^\dag_{\nk'}] = L^3 \delta_{\nk,\nk'}$, they must satisfy:
\beq\label{yknorm}
y_k g_k^* - y_k^* g_k = i
\eeq
for all $k$ at some (and hence any) time $\eta$.

The choice of the $y_k(\eta)$ corresponds to the choice of a vacuum state $|0\ket$ for the field, defined by $\ann_{\nk}{} |0\ket =0$ for all $\nk$. In the present case, as on any non stationary spacetime, it is not unique. Condition \eqref{yknorm} is not sufficient to fully determine $y_k(\eta)$. The traditional approach in inflationary models is to consider the (homogeneous and isotropic) so-called BD vacuum. In this case, the choice corresponds to the situation in which, when $k\eta_0\to -\infty$, the solution $y_k(\eta)\to\frac{1}{\sqrt{2k}}e^{-i k \eta}$; this is, the solutions are the same as the ones with positive frequencies in the flat Minkowski spacetime. In the case of inflation in a quasi-de Sitter background, this last condition together with \eqref{yknorm} correspond to fix $B_k=0$ and $|A_k|=\sqrt{\frac{\pi}{4k}}$. Readers interested in how the quasi-de Sitter case is analyzed within collapse schemes, when the BD vacuum is chosen as the initial condition (and where the prediction for the scalar spectral index is $n_s \neq 1$), are invited to see the work \cite{gabrielqdesitter}.

At this point, we must make a short digression regarding our conceptual approach and its differences with the standard picture. Any selection of a vacuum (made through the choice of the $y_k(\eta)$ that we take as positive energy modes), would be a spatially homogeneous and isotropic state of the field, as it can be seen by evaluating directly the action of a translation or rotation operators (associated with the hypersurfaces $\eta$ = constant of the background spacetime) on the state $|0\ket$. A formal proof of this can be found, for instance, in Appendix A of \cite{LLS13}. As the dynamical
evolution (through Schr\"{o}dinger equation) preserves such symmetries, the state of the system will be symmetric (homogeneous and isotropic) at all times. In fact, there is nothing, given the standard unitary evolution of the quantum theory, that could be invoked to avoid such conclusion. The issue is then: How do we account for a universe with seeds of cosmic structure, starting from an isotropic and homogeneous background spacetime and an equally symmetric vacuum state? Note that this is an open issue in all current models of inflation relying in the traditional treatment of the primordial perturbations.

As we mentioned in the Introduction, one possible solution to the aforementioned problem relies on supplementing the standard inflationary model with an hypothesis involving the modification of quantum theory so as to include a spontaneous dynamical reduction of the quantum state (sometimes referred as the self-induced collapse of the wave function) \cite{Shortcomings, PSS06}. The dynamical reduction can be considered as an actual physical process taking place independently of observers or measuring devices. Therefore, our approach regarding the origin of the primordial perturbations can be summarize as follows: a few $e$-folds after inflation has started, the universe finds itself in an homogeneous and isotropic quantum state. Then, during the inflationary regime, a quantum collapse of the wave function is triggered (by novel physics that could possibly be related to quantum gravitational effects), breaking in the process the unitary evolution of quantum mechanics and also, in general, the symmetries of the original state. That is, the post-collapse state will not be, in general, isotropic nor homogeneous. Also, the collapse mechanism functions as a generator of the metric perturbations, as will become clear below.

Readers familiar with the subject might take the posture that the problem we are characterizing is equivalent to the quantum-to-classical transition of the primordial perturbations. Several works in the literature, based on decoherence or evolution of the vacuum state into a squeezed state, have dealt with such a problem (see e.g. \cite{polarski,kiefer}). On the other hand, in Refs. \cite{PSS06,Shortcomings,LLS13} it is exposed why such arguments are not entirely convincing. Nevertheless, in the standard approach, at some point during inflation occurs the transition $\hat\Psi_{\nk} \to \Psi_{\nk} = A e^{i \alpha_{\nk}}$, with $\alpha_{\nk}$ a random phase (recall that $\Psi_{\nk}$ represents the metric perturbation). The amplitude $A$ is identified with the quantum uncertainty of $\hat \Psi_{\nk}$, i.e. $A^2=\bra 0 |\hat \Psi_{\nk}^2|0\ket$. Moreover, quantum expectation values are identified with ensemble averages of classical stochastic fields based on postulate, and the theoretical predictions agree with the observational data. Finally, note that in our approach, because our reliance on semiclassical gravity, the primordial curvature perturbation is always a classical quantity.

In the next subsections, we are going to analyze whether the replacement of the BD vacuum state by another one (motivated by different physical criteria) can affect the primordial scalar power spectrum, under the incorporation of the collapse hypothesis. As we will see, under certain conditions, one can recover a scale free spectrum for scalar perturbations, but generically there would be some characteristic deviations thereof. Note that neglecting the slow roll parameters indicates that our prediction for the primordial scalar power spectrum should be an essentially scale free spectrum, i.e. $\mathcal{P}(k)\propto \frac{1}{k^3}$. On the other hand, the observations (e.g. CMB \cite{plkinflation15}) suggest that $\mathcal{P}(k)\propto k^{-3+\mathcal{O}(\epsilon_1, \epsilon_2)}$. In other words, the scalar spectral index is such that $n_s \neq 1$. We think that when reincorporating the slow-roll parameters in the equation of motion for the field variable, we would obtain a prediction for $n_s$ consistent with the observational data. However, the modification of the $\mathcal{P}(k)$ induced by the collapse hypothesis, would be practically the same as the one obtained using the mode functions, Eq. \eqref{ykgensol}, which neglects the slow roll parameters. In fact, one is led to a similar conclusion in Refs. \cite{PSS06,pia}, in which the BD vacuum was chosen.

\subsection{Novel vacuum conditions}

As it is well known, the choice of a vacuum state is not unique in spacetimes that do not possess a time-like Killing field. This is precisely the case when, for example, we try to describe the inflationary phase of the early universe. There are several ways to choose the initial conditions; some of which can be seen in \cite{apicon2003, dani02}.

Traditionally, quantum initial conditions for perturbations in inflation are set using the BD vacuum.

The typical selection of the BD vacuum, described previously, can be deduced, for example, looking for the conditions that modes $y_k(\eta)$ must satisfy to achieve the diagonalization of the Hamiltonian of perturbations. However, those conditions are satisfied for a given initial time $\eta_0$; because as is known, in a curved spacetime the vacuum is a time-dependent notion. Hamiltonian diagonalization is the simplest approach for setting quantum initial conditions in a general spacetime, and derives the vacuum from the minimization of the Hamiltonian density. However, this approach has been criticised in the past \cite{fullingbook, fulling79}.

In order to avoid the issues raised against Hamiltonian diagonalization, the authors in \cite{handley16} motivated different initial conditions from the minimization of the \emph{renormalized} stress-energy density. The authors in \cite{handley16}, start from the action for a scalar field $\phi$ with mass $m$,
\beq\label{action2}
S=\int d^4x \sqrt{-g}\bigg[-\frac{1}{2} \nabla_a \phi \nabla_b \phi g^{ab} -\frac{1}{2} m^2\phi^2\bigg].
\eeq
By expanding the field $\phi$ in Fourier modes in the context of a FLRW spacetime as
\beq\label{modes2}
\phi(x)=\int \frac{d^3k}{(2\pi)^3 a(\eta)} \Big[\ann_{\nk} \chi_{\nk}(\eta)e^{i \nk \cdot\x}+\cre_{\nk}\chi^*_{\nk}(\eta)e^{-i \nk \cdot\x} \Big],
\eeq
if the field satisfies its equation of motion, and the commutation relation between $\ann_{\nk}$ and $\cre_{\nk}$ is the standard one, then it is well known that the mode functions $\chi_{\nk}$ must satisfy:
\barr\label{ecmods}
\chi''_{\nk}+\Big[k^2+m^2 a^2-\frac{a''}{a}\Big]\chi_{\nk}&=&0 \\
\chi_{\nk} \chi'^{*}_{\nk} - \chi_{\nk}^* \chi'_{\nk} &=& i
\label{normx}
\earr

Later, the authors computed a renormalized stress-energy tensor, $\bra 0| \hat{T}_{ab}|0 \ket_{\rm ren}$, via a Hadamard point splitting procedure. To do that, they build the stress-tensor $\bra 0| \hat{T}_{ab}|0 \ket_{\rm ren}$ using the Hadamard Green function with the mode expansion \eqref{modes2}, and subtracting off de-Witt-Schwinger geometrical terms to obtain a non-divergent quantity. Then, finally they write:
\barr\label{Trenorm}
\bra 0| \hat{T}_{00}(x)|0 \ket_{\rm ren}&=&\frac{1}{2}\int \frac{d^3k}{(2\pi)^3 a^2}(\chi'_{\nk}-\frac{a'}{a}\chi_{\nk})(\chi'^{*}_{\nk}-\frac{a'}{a}\chi^{*}_{\nk}) \nonumber \\
&+& (k^2+m^2 a^2) \chi_{\nk} \chi^{*}_{\nk}+\tilde{T}
\earr
where $\tilde{T}$ signifies additional terms arising from the renormalization process that have no dependence on the variables $\Sigma=$\{$\chi_{\nk}, \chi^{*}_{\nk}, \chi'_{\nk},\chi'^{*}_{\nk}$\}. Minimizing \eqref{Trenorm} with respect to $\Sigma$, subject to the normalization \eqref{normx}, yields the relations \cite{handley16}:
\barr\label{condone}
|\chi_{\nk}|^2 &=& \frac{1}{2\sqrt{k^2+m^2 a^2}}  \\
\chi'_{\nk}&=& \Big(-i \sqrt{k^2+m^2 a^2}+\frac{a'}{a}\Big)\chi_{\nk}
\label{condtwo}
\earr
Conditions \eqref{condone} and \eqref{condtwo} will be our guide to determine the novel vacuum conditions in the present work.

Now, let us return to our particular situation. Quantum field theory in curved spacetime describes the effects of gravity upon the quantum fields. The semiclassical Einstein equation describes how quantum fields act as the source of gravity. This equation is usually taken to be the classical Einstein equation, with the source as the quantum expectation value of the matter field stress-energy tensor operator $\hat{T}_{ab}$, that is,
\beq\label{Esemiclas}
G_{ab}=8\pi G \bra \hat{T}_{ab} \ket.
\eeq
But, this expectation value is only defined after suitable regularization and renormalization.

As already mentioned, since we are not interested in the effects on the power spectrum (and its scalar spectral index) coming from slow-roll parameters, we assume $m=0$ in Eq. \eqref{ecmods}, which is equivalent to neglect the second slow roll parameter $\epsilon_2$. Moreover, note that the equation of motion for $y_k(\eta)$, Eq. \eqref{ykmov2}, is identical to Eq. \eqref{ecmods}, which is the one obtained by the authors of \cite{handley16} when $m=0$. In particular, it involves the quantity $\frac{a''}{a}$. This contrasts with the traditional procedure involving the Mukhanov-Sasaki variable, which results in an equation of motion similar in structure to Eq. \eqref{ecmods}, but replacing $\frac{a''}{a} \to \frac{z''}{z}$, where $z\equiv \sqrt{2 \epsilon_1}  a M_P $. In other words, the quantum theory proposed by the authors of \cite{handley16} is better suited for the field $y_k(\eta)$, than for the Mukhanov-Sasaki variable because strictly if $\epsilon_1' \neq 0$, then $\frac{a''}{a} \neq \frac{z''}{z}$.

Therefore, we consider once again, the general solution \eqref{ykgensol}, which is
\beq\label{ykposta}
y_k(\eta)= A_k \Big(1-\frac{1}{k\eta}\Big) e^{-i k\eta}+B_k\Big(1+\frac{1}{k\eta}\Big) e^{i k\eta}.
\eeq
Here, without loss of generality, we will assume $A_k\in \mathbb{R}$ and  $B_k\in \mathbb{C}$. That is, only $B_k$ will carry a complex phase. Normalization \eqref{yknorm} imposes that
\beq\label{norconAB}
A_k^2-|B_k|^2=\frac{1}{2k}.
\eeq
On the other hand, identifying $\chi_{\nk}$ with $y_k(\eta)$, conditions \eqref{condone} and \eqref{condtwo} yield
\barr\label{Ak}
A_k&=&+ \sqrt{\frac{4 z_0^2+1}{8 k z_0^2}} \\
B_k&=&\frac{1}{+ \sqrt{8 k}\:|z_0|}\:e^{i \beta}
\label{Bk}
\earr
where we have defined $z_0\equiv k\eta_0$ and $\beta\equiv -2 z_0+\arctan(2 z_0)+\pi$. Note that when $z_0\to -\infty$, the BD vacuum is recovered; this is, $A_k=\frac{1}{\sqrt{2k}}$ and $B_k=0$.

Equation \eqref{ykposta} together with Eqs. \eqref{Ak} and \eqref{Bk} constitute our choice of initial vacuum conditions at time $\eta_0$. In the next subsection, we will introduce the specific collapse scheme and calculate the primordial curvature perturbation.

\subsection{Emergence of curvature perturbation within a collapse scheme}
\label{colapsos}

In this subsection, we are going to consider a modification to the standard inflationary proposal, designed to account for breaking the symmetries of the initial quantum
state, leading to the generation of the primordial inhomogeneities.

As we have claimed, when considering a quantum description for the early universe, one must face the situation in which a completely homogeneous and isotropic stage must nevertheless lead, after some time, to a universe containing actual inhomogeneities and anisotropies. This issue has been considered
at length in other works, including detailed discussions of the shortcomings of the most popular attempts to address the problem, and we will not repeat such extensive
discussions here. It is clear that such transition from a symmetric situation to one that is not, cannot be simply the result of quantum unitary evolution, since, as we noted, the dynamics does not break these initial symmetries of the system. As discussed in \cite{Shortcomings}, and despite multiple claims to the contrary (e.g. \cite{kiefer}), there is no satisfactory solution to this problem within the standard physical paradigms.

The proposal to handle this shortcoming was considered for the first time in \cite{PSS06}. There, the problem was addressed by introducing a new ingredient into the inflationary account of the origin of the seeds of cosmic structure: the self-induced collapse hypothesis. The basic idea is that an internally induced spontaneous collapse of the
wave function of the inflaton field is the mechanism by which inhomogeneities and anisotropies arise at each particular scale. That proposal was inspired on early ones
for the resolution of the measurement problem in quantum theory \cite{bohm66, Pearle76, Pearle79, pearle1989, ghirardi1985}, which regarded the collapse of the wave function as an actual physical process taking place spontaneously. Also, on the ideas by R. Penrose and L. Diosi \cite{penrose1996,diosi1987,diosi1989} who assumed that such process should be connected to quantum aspects of gravitation.

A \emph{collapse scheme} \cite{PSS06, US08} is a recipe to characterize and select the state into which each of the modes of the scalar field jumps at the corresponding time of collapse. The collapse itself is described in a purely phenomenological manner, without reference to any particular mechanism. As reported in, for instance, \cite{PSS06, DLS11, LSS12, benetti06}, the different collapse schemes generally give rise to different characteristic departures from the conventional Harrison-Zel'dovich flat primordial spectrum. There are, of course, more sophisticated theories describing the collapse dynamics, such as those in \cite{Pearle76, Pearle79, pearle1989, ghirardi1985, jmartin, CPS13, tpsingh, dasGW, LB2015}. However, we will not consider those in the present study, which is meant a first exploration of such ideas in the context of different choices of the initial quantum state.

The self-induced collapse hypothesis is based on assuming that the collapse acts similar to a ``measurement'' (in an early universe where, clearly, there are no external observers or measuring devices), this lead us to consider Hermitian operators, which in ordinary quantum mechanics are the ones susceptible of direct measurement. Therefore, we will separate $\hat y_{\nk} (\eta)$ and $\hat \pi_{\nk} (\eta)$ into their real and imaginary parts: $\hat y_{\nk} (\eta)=\hat y_{\nk}{}^R (\eta) +i\hat y_{\nk}{}^I (\eta)$ and $\hat \pi_{\nk} (\eta) =\hat \pi_{\nk}{}^R (\eta)+i \hat \pi_{\nk}{}^I (\eta)$. In this way, the operators $\hat y_{\nk}^{R, I} (\eta)$ and $\hat \pi_{\nk}^{R, I} (\eta)$ are Hermitian operators and then they can be written as,
\begin{subequations}\label{operadoresRI}
\beq
\hat{y}_{\nk}^{R,I} (\eta) = \sqrt{2} \textrm{Re}[y_k(\eta)
\hat{a}_{\nk}^{R,I}]
\eeq
\beq
\hat{\pi}_{\nk}^{R,I} (\eta) = \sqrt{2}
\textrm{Re}[g_k(\eta) \hat{a}_{\nk}^{R,I}]
\eeq
\end{subequations}
where $\hat{a}_{\nk}^R \equiv \frac{1}{\sqrt{2}}(\hat{a}_{\nk} + \hat{a}_{-\nk})$, and $\hat{a}_{\nk}^I \equiv \frac{-i}{\sqrt{2}} (\hat{a}_{\nk} - \hat{a}_{-\nk})$; and where the non-standard commutation relations for the $\hat{a}_{\nk}^{R,I}$ are,
\beq\label{creanRI}
[\hat{a}_{\nk}^{R,I},\hat{a}_{\nk'}^{R,I \dag}] = L^3 (\delta_{\nk,\nk'} \pm
\delta_{\nk,-\nk'}).
\eeq
In the last equation, the $+$ and $-$ signs correspond to the commutators with the $R$ and $I$ labels respectively; and all other commutators vanish.

Next, we will show how in our approach the quantum theory of the inflaton perturbations can be connected with the primordial curvature perturbation. Moreover, we will illustrate how the collapse process generates the seeds of cosmic structure. Here, we will proceed by choosing to work in the longitudinal gauge, and then, since $\hat{\pi}_\nk=a\hat{\dphi'}_\nk$, we will express Eq. \eqref{master} in terms of the expectation value of the conjugated momentum. Thus,
\beq\label{masterpi}
\Psi_{\nk} (\eta) \simeq  \sqrt{\frac{\epsilon_1}{2}} \frac{H}{M_P k^2}  \bra
\hat{\pi}_{\nk} (\eta) \ket .
\eeq

At the initial conformal time $\eta_0$, the state $|0\ket$ is perfectly symmetric, which implies that $ \bra \hat{\pi}_{\nk} (\eta) \ket =0$ and so, $\Psi_{\nk} =0$; i.e. there are no perturbations of the symmetric background spacetime. Afterwards, under the self-induced collapse hypothesis, at some later time $\eta_k^c$, called the time of collapse, a transition to a new state $|0\ket \to |\Theta\ket$ is produced, which does not have the initial symmetries. And in this new state, we will have that $\bra \hat{\pi}_{\nk} (\eta) \ket_\Theta \neq0$ for all $\eta\geq\eta_k^c$, and $\Psi_{\nk} \neq 0$. From Eq. \eqref{masterpi}, which was provided by the semiclassical framework, and given that all modes of the inflaton field are now in the post-collapse state $|\Theta \ket$, we can clearly see that the expectation value $\bra\hat{\pi}_{\nk} (\eta) \ket$ serves as a source for $\Psi_{\nk}$ for all $\nk $. These collapses will be assumed to take place according to certain collapse scheme which we will describe in detail below.

Taking into account Eq. \eqref{masterpi}, and that the collapse is somehow analogous to an imprecise measurement of the operators $\hat y_{\nk}^{R, I} (\eta)$ and $\hat \pi_{\nk}^{R, I} (\eta)$, our next objective is to find an equation for the dynamics of the expectation values of $\bra  \hat{\pi}^{R, I}_{\nk} (\eta)   \ket$, evaluated in the post-collapse state. This equations, as we shall see, will be related to the values $\bra  \hat{y}^{R,I}_{\nk} (\tc) \ket$ and $\bra  \hat{\pi}^{R, I}_{\nk} (\tc) \ket$, through the proposed collapse scheme.

In the vacuum state, $\hat{y}_{\nk}$ and $\hat{\pi}_{\nk}$ individually are distributed according to Gaussian wave functions centered at zero with spread $(\Delta \hat{y}_{\nk})^2_0$ and $(\Delta\hat{\pi}_{\nk})^2_0$, respectively. Our assumption is that the effect of the collapse on a state is analogous to some sort of approximate measurement. Therefore, after the collapse the expectation values of the field and momentum operators, in each mode, will be related to the uncertainties of the initial state.

We will adopt a collapse scheme, where it is assumed that the expectation values of the field mode $\hat{y}^{R,I}_{\nk}$ and their conjugate momentum $\hat{\pi}^{R,I}_{\nk}$ acquire independent values randomly, and where the expectation (in the new state $|\Theta\ket$) at the time of collapse is given by:
\begin{subequations}\label{esquemaind}
\beq
\bra \hat{y}^{R,I}_{\nk}(\tc)\ket_\Theta  = \lambda_1\:x_{\nk,1}^{R,I}
  \sqrt{\left(\Delta \hat{y}^{R,I}_{\nk} (\tc) \right)^2_0}
  \eeq
  \beq
  \bra \hat{\pi}^{R,I}_{\nk}(\tc) \ket_\Theta = \lambda_2\:x_{\nk,2}^{R,I}
  \sqrt{\left(\Delta \hat{\pi}^{R,I}_{\nk} (\tc) \right)^2_0}
  \eeq
\end{subequations}
The parameters $\lambda_1$ and $\lambda_2$ are viewed as ``switch-off/on'' parameters. This is, they can only take the values 0 or 1 depending on which variable
$\hat{y}^{R,I}_{\nk}$, $\hat{\pi}^{R,I}_{\nk}$ or both is affected by the collapse. For instance, in past works \cite{PSS06, Leon11}, the name \emph{independent
scheme} was coined for the case $\lambda_1=1=\lambda_2$, i.e.  $\hat{y}^{R,I}_{\nk}$ and $\hat{\pi}^{R,I}_{\nk}$ are both affected independently by the collapse. In this scheme the expectation value jumps to a random value $x_{\nk}^{(R,I)}$ multiplied by the uncertainty of the vacuum state of the field. The random variables $x_{\nk,1}^{(R)}, x_{\nk,1}^{(I)}, x_{\nk,2}^{(R)}$ and $x_{\nk,2}^{(I)}$ are selected from a Gaussian distribution centered at zero, with unity spread, and all of them are assumed statistically uncorrelated. In Appendix \ref{appA} we show the quantum uncertainties $\left(\Delta\hat{y}^{R,I}_{\nk} (\tc)\right)^2_0$ and $\left(\Delta\hat{\pi}^{R,I}_{\nk} (\tc) \right)^2_0$ in the vacuum state within this collapse scheme.

Having established the relation between the curvature perturbation and the quantum matter fields [see Eq. \eqref{masterpi}], as well as the characterization of the collapse, the next aim is to present an explicit expression for the curvature perturbation in terms of the parameters characterizing the collapse scheme. In order to achieve that goal, we must first find an expression for the evolution of the expectation values of the fields. In fact, as can be seen from Eq. \eqref{masterpi}, we will only be concerned with the expectation value of the conjugated momentum $\bra \hat{\pi}_{\nk} (\eta)\ket$.

In Appendix \ref{appA}, we show that:
\barr\label{expecpitexto}
\bra \hat{\pi}_{\nk}(\eta) \ket_\Theta &=& -L^{3/2}\Big[F(k\eta,z_k) \:|y_k(\tc)|\:\lambda_1 \:X_{\nk,1} +\nn
&+& G(k\eta,z_k) \:|g_k(\tc)|\:\lambda_2\: X_{\nk,2} \Big]
\earr
where $X_{\nk,1}\equiv x_{\nk,1}^R+i\: x_{\nk,1}^I$, $X_{\nk,2}\equiv x_{\nk,2}^R+i \: x_{\nk,2}^I$, and the functions $F(k\eta,z_k)$, $G(k\eta, z_k)$, $|y_k(\tc)|$ and $|g_k(\tc)|$ are also defined in Appendix \ref{appA}. Notice that the constants $A_k$ and $B_k$ appear in all these functions, and is in such constants that the information about the initial conditions is found, through its dependence with $z_0\equiv k\eta_0$. Also, the parameter $z_k$ is defined as $z_k \equiv k\tc$; thus, $z_k$ is directly associated to the time of collapse $\tc$.

Finally, substituting Eq. \eqref{expecpitexto} in Eq. \eqref{masterpi}, we find the expression for the curvature perturbation (in the longitudinal gauge):
\barr\label{masterpsi}
\Psi_{\nk} (\eta) &=&  -\sqrt{\frac{\epsilon_1}{2}} \frac{H\:L^{3/2}}{M_P\: k^2}\Big[F(k\eta,z_k) \:|y_k(\tc)|\:\lambda_1 \:X_{\nk,1} + \nn
&+& G(k\eta,z_k) \:|g_k(\tc)|\:\lambda_2\: X_{\nk,2} \Big]
\earr

The curvature perturbation $\Psi$ in the longitudinal gauge, is a constant quantity for modes ``outside the horizon'' during any given cosmological epoch, but not during
the transition between epochs. In fact, during the transition from the inflationary stage to the radiation dominated stage, $\Psi$ is amplified by a factor of $1/\epsilon_1$ \cite{brandenberger1993,Deruelle1995}. On the other hand, an useful gauge-invariant quantity often encountered in the literature is the variable $\mathcal{R} (x)$. The field $\mathcal{R}(x)$ is a field representing the curvature perturbation in the \emph{comoving gauge}. Its Fourier transform, represented by $\mR_{\nk}$, is constant for modes greater than the Hubble radius (irrespectively of the cosmological epoch), i.e. for modes with $k \ll \mH = aH$ (and assuming adiabatic perturbations). This is, the value of $\mR_{\nk}$ during inflation (in the limit $k \ll \mH$) will remain unchanged during the post-inflationary evolution, until the mode ``re-enters the horizon'', namely when $k\simeq \mH$, at the later radiation/matter dominatted epochs. The curvature perturbation in the comoving gauge $\mathcal{R}$ and the curvature perturbation in the longitudinal gauge $\Psi$ are related by $\mathcal{R} \equiv \Psi + (2/3)(\mH^{-1} {\Psi}' + \Psi)/(1+\omega)$, with $\omega \equiv p/\rho$. Therefore, for modes such that $k \ll \mH$, during the inflationary epoch $\omega+1 \simeq 2 \epsilon_1/3$, it is found that
\beq\label{R}
\lim_{ k \ll \mH} \mathcal{R}_{\nk}(\eta) \simeq \lim_{ k \ll \mH} \frac{\Psi_{\nk} (\eta)}{\epsilon_1}
\eeq
with $\Psi_{\nk} (\eta)$, calculated during inflation, in the limit such that the modes are well outside the horizon.

In the next subsection, we are going to find an expression for the scalar power spectrum of the perturbations $\mathcal{R}_{\nk}$, namely $P(k)$. The power spectrum serves as the initial condition to obtain the angular spectrum, denoted usually in the literature as $C_l$, which is the theoretical prediction that is contrasted with the observations directly. Henceforth, we will obtain an expression for $\mathcal{R}_{\nk}$ during inflation explicitly, for the observationally relevant modes. Specifically, we take the limit $|k\eta|\to 0$ in the functions $F(k\eta,z_k)$ and $G(k\eta,z_k)$, and we define $M(z_0, z_k)\equiv \lim_{|k\eta|\to 0} F(k\eta, z_k)\:|y_k(\tc)|$ and $N(z_0, z_k)\equiv \lim_{|k\eta|\to 0} G(k\eta, z_k) \:|g_k(\tc)|$. Thus,
\beq\label{Rkposta}
\mR_{\nk} = \frac{-HL^{3/2}}{\sqrt{2\epsilon_1}M_P k^2} \Big[M(z_0, z_k)\:\lambda_1 \:X_{\nk,1}+N(z_0, z_k)\:\lambda_2\: X_{\nk,2} \Big]
\eeq
Note the explicit dependence on the initial time $\eta_0$ through the quantity $z_0 \equiv k\eta_0$. Equation \eqref{Rkposta} is the main result of this section.

We strongly remark that the random variables $X_{\nk}$ corresponding to the collapse scheme are fixed after the collapse of the wave function has occurred. In other words, if we somehow knew their exact value, we would be able to predict the exact value for $\mR_{\nk}$. We will do make use of the statistical properties of these random variables to be able to make theoretical predictions for the observational quantities. For a more detailed comparison between our approach and the traditional inflationary picture see Ref. \cite{LSS12}.

Next, we will consider whether, and under what circumstances, one can obtain a prediction for the power spectrum of the scalar perturbations $\mathcal{R}_{\nk}$, for the case of the observationally relevant modes, which have wavelengths greater than the Hubble radius at the time of inflation.

\subsection{Primordial scalar power spectrum}
\label{espectro}

In this subsection, we will calculate the primordial power spectrum of the scalar perturbations $\mR_{\nk}$, and analyze its relation with the CMB observations under our approach.

The temperature anisotropies of the CMB, $\delta T/T_0$, are the most direct observational quantity available, with $T_0$ the mean temperature today. Expanding $\delta T/T_0$ using spherical harmonics, the coefficients $a_{lm}$ are
\beq\label{alm0}
a_{lm} = \int \Theta (\hat n) Y_{lm}^{*} (\theta,\varphi) d\Omega,
 \eeq
with $\hat n = (\sin \theta \sin \varphi, \sin \theta \cos \varphi, \cos\theta)$ and $\theta,\varphi$ the coordinates on the celestial two-sphere. By defining $\Theta (\hat {n}) \equiv \delta T (\hat n)/ T_0$ and assuming instantaneous recombination, the relation between the primordial perturbations and the observed CMB temperature anisotropies is
\beq\label{mastertemp}
\Theta (\hat n) = [\Psi + \frac{1}{4} \delta_\gamma] (\eta_D) + \hat n \cdot
\vec{v}_\gamma (\eta_D) + 2 \int_{\eta_D}^{\eta_0} \Psi'(\eta) d\eta,
\eeq
where $\eta_D$ is the time of decoupling; $\delta_\gamma$ and $\vec{v}_\gamma$ are the density perturbations and velocity of the radiation fluid.

On the other hand, the temperature anisotropies in Fourier modes is
\beq
\Theta (\hat n) = \sum_{\nk} \frac{\Theta (\nk)}{L^3} e^{i \nk \cdot R_D \hat
n}
\eeq
being $R_D$ the radius of the last scattering surface. Then, the fluid motion equations can be solved with the initial condition provided by the curvature perturbation during inflation. Furthermore, using that $e^{i \nk \cdot R_D \hat n} = 4 \pi  \sum_{lm} i^l j_l (kR_D) Y_{lm} (\theta,\varphi) Y_{lm}^{*} (\hat k )$,  expression \eqref{alm0} can be rewritten as
\beq\label{alm1}
a_{lm} = \frac{4 \pi i^l}{L^3} \sum_{\nk} j_l (kR_D) Y_{lm}^{*}(\hat k)
\Theta(\nk),
\eeq
with $j_l (kR_D)$ the spherical Bessel function of order $l$. To incorporate the linear evolution that relates the initial curvature perturbation $\mR_{\nk}$ and the anisotropies $\Theta (\nk)$, is defined a transfer function $T(k)$. This function results of solving the fluid motion equations (for each mode) with the initial condition provided by the curvature perturbation $\mR_{\nk}$, and then make use of Eq. \eqref{mastertemp} to relate it with the temperature anisotropies. In this way, $\Theta(\nk) = T(k) \mR_{\nk}$. Therefore, the coefficients $a_{lm}$, in terms of the modes $\mR_{\nk}$, are given by
\beq\label{alm2}
a_{lm} = \frac{4 \pi i^l}{ L^3} \sum_{\nk} j_l (kR_D) Y_{lm}^{*}(\hat k) T(k)\mR_{\nk}
\eeq
with $\mR_{\nk}$ during inflation, and in the limit $k \ll \mH$.

Now, substituting the explicit form of $\mR_{\nk}$ given by Eq. \eqref{Rkposta} in Eq. \eqref{alm2}, the coefficients $a_{lm}$ are directly related to the random variables $X_{\nk}$. Notice that the coefficients $a_{lm}$ are a sum of random complex numbers (i.e. a sum over $\nk$), where each term is characterized by the random complex variables $X_{\nk}$. This leads to what can be considered effectively as a two-dimensional random walk. As it is well known, one cannot give a perfect estimate for the direction of the final displacement resulting from the random walk. However, it is possible to give an estimate for the length of the total displacement. In the present case, such length can be naturally associated with the magnitude $|a_{lm}|^2$. Hence, the most likely value of $|a_{lm}|^2$ can be estimated, and thus interpret it as our theoretical prediction for the observed $|a_{lm}|^2$. Moreover, since the collapse is being modeled by a random process, we can consider a set of possible realizations of such a process characterizing the universe in an unique manner, i.e., through the random variables $X_{\nk}$. If the probability distribution function of $X_{\nk}$ is Gaussian, then we can identify the most likely value $|a_{lm}|^2_{\text{ML}}$ with the mean value $\overline{|a_{lm}|^2}$ of all possible realizations. This is, $|a_{lm}|^2_{\text{ML}}= \overline{|a_{lm}|^2}$. The most likely value $|a_{lm}|^2_{\text{ML}}$ in our collapse scheme is explicitly given in Appendix \ref{appB}.

Since we are assuming that the $x_{\nk}^{R,I}$ variables are uncorrelated, the ensemble average of the product of these random variables satisfies
\beq\label{distribucionx}
\overline{x^R_{\nk} x^R_{\nk'}} = \delta_{\nk,\nk'} + \delta_{\nk,-\nk'}
\quad \overline{x^I_{\nk} x^I_{\nk'}} = \delta_{\nk,\nk'} -
\delta_{\nk,-\nk'}
\eeq
We have also considered the correlation between the modes $\nk$ and $-\nk$ in accordance with the commutation relation given by $[\hat{a}^{R}_{\nk},\hat{a}^{R\dag}_{\nk'}]$ and $[\hat{a}^{I}_{\nk},\hat{a}^{I \dag}_{\nk'}]$.

Typically, the observational CMB data is presented in terms of the angular power spectrum,  $C_l$. The definition of $C_l$ is given in terms of the coefficients $a_{lm}$ as $C_l=(2l+1)^{-1} \sum_m |a_{lm}|^2$. Therefore, we can use the prediction for $|a_{lm}|^2_{\text{ML}}$ for our collapse scheme considered, and give a theoretical prediction for $C_l$. Thus, form Eqs. \eqref{distribucionx} and using our values for the $|a_{lm}|^2_{\text{ML}}$ we can write,
\beq\label{clcolapso}
C_l = { 4 \pi} \int_0^\infty \frac{dk}{k} j_l^2 (kR_D) T(k)^2 A\:Q(z_0, z_k)
\eeq
where the explicit form of the function $Q(z_0, z_k)$ is shown in Appendix \ref{appB}, and $A$ is:
\beq\label{Amp}
A=\frac{H^2}{2\pi^2 M_P^2\epsilon_1}
\eeq
Also, we have taken the limit $L\to \infty$ and $\nk \to$ continuum in order to go from sums over discrete $\nk$ to integrals over $\nk$.

In the standard inflationary paradigm, a well-known result is that the dimensionless power spectrum $P (k)$ for the perturbation $\mR_{\nk}$ and the $C_l$ are related by
\beq\label{cl}
C_l = {4 \pi} \int_0^\infty \frac{dk}{k}  j_l^2 (kR_D) T(k)^2 P(k).
\eeq
Therefore, by comparing Eq. \eqref{clcolapso} with Eq. \eqref{cl} we can extract an equivalent power spectrum\footnote{Bear in mind that there are two power spectrum in the literature: the dimensional power spectrum $\mathcal{P} (k)$ and the dimensionless power spectrum $P(k)$. The latter is defined in terms of the former by $ P(k) \equiv  (k^3/2\pi^2)\mathcal{P}(k)$. We are expressing our main result as the $P(k)$}, which finally turns out to be:
\beq\label{pscolapso}
P(k) = \frac{H^2}{2\pi^2 M_P^2\epsilon_1} Q (z_0, z_k)
\eeq
Equation \eqref{pscolapso} is the main result of this work. Notice that because of $z_k=k\tc$ and $z_0=k\eta_0$, the function $Q (z_0, z_k)$ depends on $k$ explicitly.

In the next section, we will discuss the results, compare them with previous works, and analyze under which conditions one can recover an essentially scale free spectrum of primordial inhomogeneities, as suggested by the observations.

We would like to end this section by making some comments about our prediction for the power spectrum. Our model gives a direct theoretical prediction for the observed $C_l$, Eq. \eqref{clcolapso}, and then from such expression we have read what can be identified as the ``power spectrum'' in the traditional approach of inflation. However, note that this is conceptually different from the traditional approach \cite{Mukhanov2005} in which the power spectrum is obtained from the two-point correlation function $\bra 0 | \hat \mR_{\nk} \hat \mR_{\nk'}^{*} | 0\ket$. In contrast, our power spectrum is obtained from $\overline{\bra\hat{\pi}_{\nk} \ket \bra \hat{\pi}_{\nk '}\ket^{*}}$, where the expectation values are evaluated at the post-collapse state. In Appendix \ref{appC}, we show in detail the calculation of the CMB temperature angular spectrum and its relation with the scalar power spectrum. The interested reader can find there an explanation on how to calculate the power spectrum within the collapse framework, and why our proposal does not rely on the quantum two-point correlation function.

\section{Results and discussion}
\label{seccuatro}

Let us summarize briefly the results obtained in the present manuscript. We started by choosing a novel initial quantum vacuum state for the perturbations of the inflaton [whose mathematical description is given by Eqs. \eqref{ykposta}, \eqref{Ak} and \eqref{Bk}]. Then, we included the collapse hypothesis and finally arrived at Eq. \eqref{pscolapso} for the primordial scalar power spectrum. Note that, as already mentioned, the vacuum $y_k(\eta)$ in Eq. \eqref{ykposta} includes the initial condition of the BD vacuum if $z_0\to -\infty$. However, notice that the physical criteria for the choice of both vacuum states are very different.

Now, from our result shown in Eq. \eqref{pscolapso} for $P(k)$, we are going to analyze under which conditions a scale free spectrum can be obtained (this is, when the function $Q(z_0, z_k)$ does not depend on $k$ and results in a constant). Also, we will analyze those cases where $P(k)$ shows small deviations from a scale invariant spectrum, but are still consistent with observational data. Here, we will adopt $\lambda_1=1=\lambda_2$ (i.e. the independent collapse scheme).

We consider three cases, according to $|z_0|$ values:
\subsection{$|z_0|\to 0$}
\label{zozero}

In this case, the (dimensional) scalar power spectrum results in $\mathcal{P}(k)\propto \frac{1}{k^5}$, both in the standard scenario and in ours having included the hypothesis of collapse. In our approach, if $|z_0|\to 0$, then the time of collapse must satisfy $|z_k| \to 0$ since the collapse always occurs for $\tc>\eta_0$. There is no possible parametrization for $\tc$, such that the spectrum is scale invariant; the proof can be seen in the Appendix \ref{appD}. We stress that, the loss of scale invariance in the resulting power spectrum, is  not due to the physics behind the collapses, but because the novel vacuum choice is not the best option for smaller values of $|z_0|$.

\subsection{$|z_0|\to \infty$}
\label{zoinf}

In this case, we observe that taking the limit $|z_0|\to \infty$, implies that the function $Q(z_0, z_k)$ results in exactly the same function as the one shown in Eq. (88) of Ref. \cite{PSS06}, i.e.
\begin{equation}\label{C(k)dePSS}
Q(z_0, z_k)\to C(z_k) = \frac{1}{8}\Big[1 + \frac{ 2 }{ z_k^{2} }\sin^{2}(z_k) - \frac{ 1 }{ z_k }\sin(2z_k)\Big]
\end{equation}
Note that the aforementioned result of Ref. \cite{PSS06} was obtained using the BD vacuum state. Equation \eqref{C(k)dePSS} is expected because if $|z_0|\to \infty$, the initial condition of the BD vacuum is recovered.

In Ref. \cite{PSS06}, it was shown that if $|z_k|\to\infty$ or $|z_k|\to 0$ then the function $C(z_k)$ is a constant. Thus, leading to an exactly scale invariant power spectrum. Also, if $\tc=\frac{\mathcal{A}}{k}$ is assumed (with $\mathcal{A}$ a constant), whatever the value of $z_k$, then the resulting spectrum is scale free, and its observational analysis can be consulted, for instance, in Refs. \cite{LSS12, benetti06}.

\subsection{Other $|z_0|$ cases}
\label{zointerm}

\begin{figure}[h!]
\begin{center}
\includegraphics[width=9cm,angle=0]{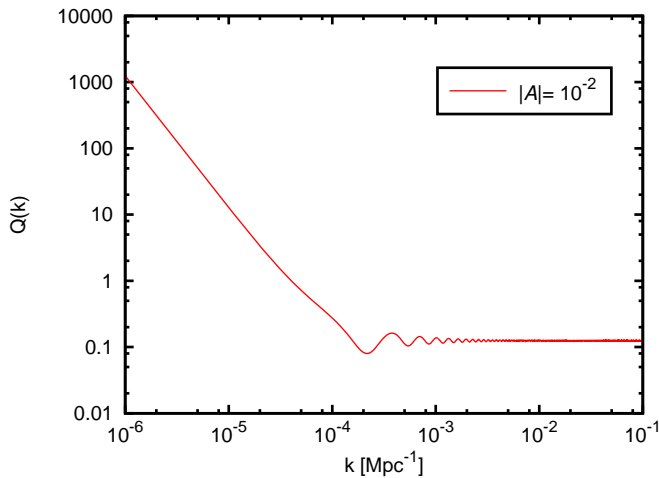}
\end{center}
\caption{The function $Q(k)$ vs. $k$, for intermediate $|z_0|$ values and when the parametrization $\tc=\frac{\mathcal{A}}{k}$ is assumed. The values considered are$ |\mathcal{A}|=10^{-2}$ and $|\eta_0|=10^4 {\rm Mpc}$. The behavior of $Q$ is nearly constant except for large scales (i.e. small $k$ values), for which, small deviations from a scale free spectrum are expected in our prediction for the $C_l$.} \label{Fig1}
\end{figure}

For intermediate $|z_0|$ values, that is, values not included in the cases A and B described previously, in Fig. \ref{Fig1} we plot $Q(k)$ vs. $k$ having chosen the time of collapse as $\tc=\frac{\mathcal{A}}{k}$ ($\mathcal{A}$ a constant), with $|\mathcal{A}|=10^{-2}$ and $|\eta_0|=10^4\: {\rm Mpc}$. Note that when a parametrization for $\tc$ is chosen and the time $\eta_0$ is set, the function $Q(k)$ is only dependent on $k$. As it can be seen, the resulting function $Q(k)$ is constant for large values of $k$, while departures from a constant behavior for lower $k$. Therefore, we expect that the (dimensional) scalar power spectrum results in $\mathcal{P}(k)\sim\frac{1}{k^3}$ for large $k$ values, while departures from the standard prediction affect the smallest ones. Since the observational relevant modes are such that $k\in[10^{-6}, 10^{-1}]$ ${\rm Mpc^{-1}}$, only for observationally relevant small $k$ values (i.e. low multipoles $l$), a difference is expected between our prediction for the $C_l$ and that corresponding to a perfectly scale invariant spectrum. Also, note that Fig. \ref{Fig1} includes a wide range of intermediate $|z_0|$ values and, in addition, the graph is representative for $|z_k|\to 0$ and intermediate $|z_k|$ values.

\begin{figure}[h!]
\begin{center}
\includegraphics[width=9cm,angle=0]{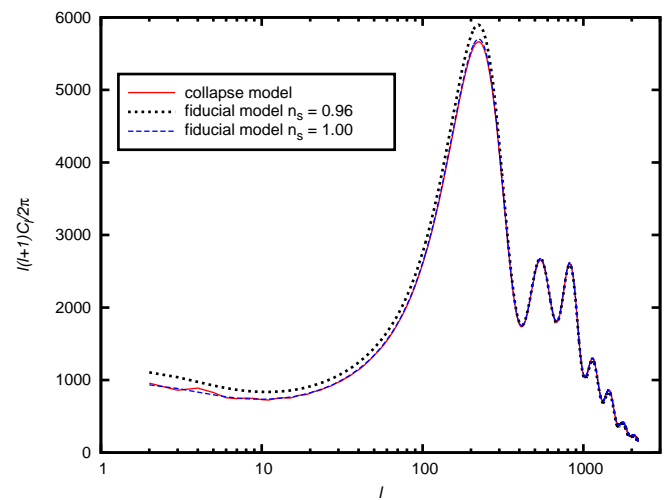}
\end{center}
\caption{Our prediction for the CMB angular spectrum within an inflationary collapse model, with $|z_k|=|\mathcal{A}|=10^{-2}$, and having chosen a different vacuum state than the traditional one, being $|\eta_0|=10^4 {\rm Mpc}$. As a reference, we also show a fiducial model coming from the best-fit to the Planck data \cite{plkinflation15}, with scalar spectral index $n_s=0.96$ (dotted line) and $n_s=1$ (dashed line). See the text for details.} \label{Fig2}
\end{figure}

Although here we will not perform a complete statistical analysis with the observational data, in Fig. \ref{Fig2} we show our predicted angular spectrum $C_l$ and compare it with a fiducial model.

In order to perform our analysis, we modified the public available CAMB code \cite{CAMB}. The cosmological parameters of our fiducial flat $\Lambda$CDM model considered are: baryon density in units of the critical density $\Omega_{{\rm b}}h^{2}=0.02225$, dark matter density in units of the critical density $\Omega_{{\rm cdm}}h^2=0.1198$, Hubble constant $H_0=67.27\: {\rm km\:s^{-1}Mpc^{-1}}$, reionization optical depth $\tau=0.079$, and the scalar spectral index, $n_s=0.96$. Those are the best-fit values presented by the Planck Collaboration \cite{plkinflation15}. Recall that, we have neglected the effects on the power spectrum and the scalar spectral index coming from the slow-roll parameters. Consequently, our angular spectrum should be compared with a canonical scale free spectrum.

In Fig. \ref{Fig2}, we present three plots: One, is the fiducial model described previously. Another one is a quasi fiducial model with the best-fit values from Planck, except for the spectral index, for which $n_s=1$. And, the remaining plot, corresponds to the predicted curve in our model, also with $n_s=1$.

As it can be seen, our prediction agrees very well with the standard prediction curve plotted with the best-fit values from the Planck data. As anticipated, only small differences for low multipole values appear, where the cosmic variance is dominant.

We conclude this section with a final remark: the fact of having chosen a vacuum different from that which is typically chosen as the initial condition for inflation, led us to a function $Q(z_0, z_k)$ that is generically different from that of the authors in \cite{PSS06}. Then, one could think that it would be very difficult to find a parametrization for $\tc$, such that the shape of the power spectrum would be compatible with the CMB observations. However, as inferred from the plots, the parametrization $\tc=\frac{\mathcal{A}}{k}$ is still valid for a wide range of $\eta_0$ values.

\section{Conclusions}
\label{conclusions}

In this work, we have analyzed under which conditions one can recover an essentially scale free spectrum of primordial inhomogeneities, when the standard BD vacuum is replaced by another one that minimizes the renormalized stress-energy tensor via a Hadamard procedure. This new prescription for selecting the initial vacuum state is better suited for cosmological models built to give a description of the early universe, particularly those that include the self-induced collapse proposal within the semiclassical gravity picture.

We found that a scale invariant scalar power spectrum can be obtained (and compatible with the CMB observations), for a wide range of initial times $\eta_0$. By choosing for the collapse time the parametrization $\tc=\mathcal{A}/k$ (being $\mathcal{A}$ a constant), in Fig. \ref{Fig2}, we show that our predicted angular spectrum  $C_l$ agrees very well with the standard prediction curve plotted with the best-fit values from the recent Planck data. As anticipated in our analysis, only small differences for low multipole $l$ values appear, where the cosmic variance is dominant.

For $|z_0|=k|\eta_0|\to 0$ the choice of this novel vacuum does not lead to a scale invariant scalar power spectrum, but not due to the collapse hypothesis. As a matter of fact, this is generically true for the standard inflationary model using the new vacuum choice. In other words, small values of $|k\eta_0|$ are not allowed by the observations using the new vacuum state either within the standard inflation or using the self-induced collapse hypothesis.

On the other hand, for values $|z_0| \gtrsim 1$, it is possible to obtain a scale free spectrum concordant with observations. In particular, for $|z_0| \gg 1$, the initial conditions are the same as the one provided by the BD vacuum.

The fact of having chosen a vacuum different from that which is typically chosen as the initial condition for inflation, led us to the function $Q(z_0, z_k)$, shown explicitly in Eq. \eqref{Qposta}. The obtained function is generically different from that of the authors in \cite{PSS06}, who also considered the collapse proposal but using the standard BD vacuum. Therefore, one could think that it would be very difficult to find a parametrization for $\tc$, such that the power spectrum would become scale free, and, as a consequence, compatible with the CMB data. However, we have found that the parametrization $\tc=\frac{\mathcal{A}}{k}$, which is the same as the one originally proposed in all the previous works based on the collapse hypothesis, is still valid for a wide range of $\eta_0$ values. Thus, we conclude that the collapse mechanism might be of a more fundamental character than previously suspected.

Note that the model considered here involved some phenomenological characterization of the self-induced collapse proposal. For instance, the dependence of the time of collapse on each mode $k$, and the Gaussian distribution in the random variables. However, this characterization can be taken as ansatz modifications of the standard inflationary scenario, inspired by collapse schemes proposals based on spontaneous individual collapses (e.g. the GRW model \cite{ghirardi1985}). The GRW objective reduction model has been originally proposed to deal with the quantum measurement problem, independent of any cosmological context. In this work, we used a GRW-inspired collapse scheme, by incorporating some of its generic features. Nevertheless, we think that a dynamical reduction mechanism (which can be seen as less \emph{ad hoc} than the one considered here), such as the Continuous Spontaneous Localization (CSL) model \cite{pearle1989,bassi2003}, can be subjected to the same analysis presented in the present paper.

Finally, it is also important to mention some previous results regarding other observables; specifically, the primordial bispectrum \cite{LSS12,bispec} and tensor modes \cite{lucila,abhishek}. Those results were obtained under our self-induced collapse proposal but maintaining the usual choice of the BD vacuum. In respect to the bispectrum, we have obtained a completely different shape than the usual one. As a matter of fact, the characterization is not based on the usual quantum three-point function. Meanwhile, a possible detection of primordial gravitational waves would be considered as the ``smoking-gun'' between our proposal, based on semiclassical gravity, and the traditional one. Our framework predicts a strong suppression of the tensor modes amplitude; essentially undetectable by any present or future experiments. Based on the results obtained in this paper, we think that the aforementioned predictions will remain unchanged.

\begin{acknowledgements}
G.R.B. is supported by CONICET (Argentina). G.R.B. acknowledges
support from the PIP 112-2012-0100540 of CONICET (Argentina). G.L. acknowledges financial support from CONICET (Argentina). We thank the anonymous referee for her/his comments and objective suggestions that contributed to the clarity of the work.
\end{acknowledgements}

\appendix
\section{Explicit equations of Sect. \ref{colapsos}}\label{appA}

In this Appendix, we are going to show some steps to arrive at Eq. \eqref{expecpitexto}. We will use the collapse scheme chosen in Eq. \eqref{esquemaind}, where appear the expectation values of $\bra  \hat{y}^{R,I}_{\nk} (\tc) \ket_\Theta$ and $\bra  \hat{\pi}^{R, I}_{\nk} (\tc) \ket_\Theta$, evaluated at the collapse time, and related to the quantum uncertainties of the vacuum state.

From Eqs. \eqref{operadoresRI}, together with definitions of $\hat{a}_{\nk}^R$ and $\hat{a}_{\nk}^I$ and the non-standard commutation relations for the $\hat{a}_{\nk}^{R,I}$ in Eq. \eqref{creanRI}, we can rewrite the quantum uncertainties of the vacuum state $y_k(\eta)$ of Eq. \eqref{ykposta} (at collapse time) as
\barr
\left(\Delta \hat{y}^{R,I}_{\nk} (\tc) \right)^2_0 &=& \frac{L^3}{4}
|y_k(\tc)|^2 \nonumber \\
\left(\Delta \hat{\pi}^{R,I}_{\nk} (\tc) \right)^2_0 &=& \frac{L^3}{4}
|g_k(\tc)|^2 \nonumber
\earr
where $g_k = y_k'-\mH y_k$. Then, the collapse scheme can be written as:
\begin{subequations}\label{esqapp}
\beq
\bra \hat{y}^{R,I}_{\nk}(\tc)\ket_\Theta  = \lambda_1\:x_{\nk,1}^{R,I}\: \frac{L^{\frac{3}{2}}}{2} |y_k(\tc)|
    \eeq
  \beq
  \bra \hat{\pi}^{R,I}_{\nk}(\tc) \ket_\Theta = \lambda_2\:x_{\nk,2}^{R,I}\:\frac{L^{\frac{3}{2}}}{2} |g_k(\tc)|
  \eeq
\end{subequations}

Now, we need the values of $\hat{y}_{\nk}(\eta)$ and $\hat{\pi}_{\nk}(\eta)$ for $\eta \geq \tc$, in the post-collapse state. To do this, we introduce the quantity
$d^{R,I}_{\nk} \equiv \langle\Theta|\ann_{\nk}^{R,I}|\Theta\rangle$, that determines the expectation value of the field and momentum operator for the mode  $\nk$ at all
times after the collapse. That is, from Eq. \eqref{operadoresRI}, we have
\barr\label{expeceta1}
\bra \hat{y}_{\nk}^{R,I} (\eta) \ket_{\Theta} = \sqrt{2}\textrm{Re}[y_k(\eta) d_{\nk}^{R,I}]\\
\bra \hat{\pi}_{\nk}^{R,I} (\eta) \ket_{\Theta} = \sqrt{2}\textrm{Re}[g_k(\eta) d_{\nk}^{R,I}]
\label{expeceta}
\earr
which corresponds to expectation values at any time after the collapse in the post-collapse state $| \Theta \ket$. One can then relate the value of $d_{\nk}^{R,I}$ with the value of the expectation value of the fields operators at the time of collapse $\bra \hat{y}_{\nk}^{R,I} (\tc) \ket_{\Theta} = \sqrt{2} \textrm{Re}[y_k(\tc) d_{\nk}^{R,I}]$,  $\bra
\hat{\pi}_{\nk}^{R,I} (\tc) \ket_{\Theta} = \sqrt{2} \textrm{Re}[g_k(\tc) d_{\nk}^{R,I}]$. Using the latter relations to express $d_{\nk}^{R,I}$ in terms of the
expectation values at the time of collapse, and substituting it in Eq. \eqref{expeceta}, we obtain an expression for the expectation value of the momentum field operator in terms of the expectation value at the time of collapse. Since $\bra \hat{\pi}_{\nk}(\eta) \ket_\Theta=\bra \hat{\pi}_{\nk}^{R} (\eta) \ket_\Theta+i\:\bra \hat{\pi}_{\nk}^{I} (\eta) \ket_\Theta$, we write
\barr\label{preexpecpi}
\bra \hat{\pi}_{\nk} (\eta) \ket_\Theta &=& \frac{1}{D(z_k)}\bigg\{F(k\eta,z_k)  \Big[\bra
\hat{y}_{\nk}^{R} (\tc) \ket_\Theta +i \bra\hat{y}_{\nk}^{I} (\tc) \ket_\Theta \Big]+\nn
&+& G(k\eta,z_k) \Big[ \bra \hat{\pi}_{\nk}^{R} (\tc) \ket_\Theta +i \bra\hat{\pi}_{\nk}^{I} (\tc) \ket_\Theta \Big] \bigg\}
\earr
where $z_k\equiv k\tc$. On the other hand, functions $D$, $F$ and $G$ are defined by:
\barr\nn
D(z_k)&\equiv& \textrm{Im}[g_k(\tc)] \textrm{Re}[y_k(\tc)]-\textrm{Im}[y_k(\tc)] \textrm{Re}[g_k(\tc)]\nn
F(k\eta,z_k)&\equiv& \textrm{Re}[g_k(\eta)] \textrm{Im}[g_k(\tc)]-\textrm{Im}[g_k(\eta)] \textrm{Re}[g_k(\tc)] \nn
G(k\eta,z_k)&\equiv& \textrm{Im}[g_k(\eta)] \textrm{Re}[y_k(\tc)]-\textrm{Re}[g_k(\eta)] \textrm{Im}[y_k(\tc)]\nn
\earr
Substituting in Eq. \eqref{preexpecpi} what was found in Eqs. \eqref{esqapp}, we can finally write,
\barr\label{expecpi}
\bra \hat{\pi}_{\nk}(\eta) \ket_\Theta &=& \frac{L^\frac{3}{2}}{2D(z_k)}\Big[F(k\eta,z_k) \:|y_k(\tc)|\:\lambda_1 \:x_{\nk,1} +\nn
&+& G(k\eta,z_k) \:|g_k(\tc)|\:\lambda_2\: x_{\nk,2} \Big]
\earr
with $X_{\nk,1}\equiv x_{\nk,1}^R+i\: x_{\nk,1}^I$ and $X_{\nk,2}\equiv x_{\nk,2}^R+i \: x_{\nk,2}^I$. From the equation for $y_k(\eta)$, Eq. \eqref{ykposta} at $\tc$, together with the relation between $A_k$ and $B_k$ given by Eq. \eqref{norconAB}, we obtain that $D=-1/2$. The explicit forms for $F$ and $G$ functions turn out to be:
\barr\label{fg}
F(k\eta,z_k)&=& (A_k^2-|B_k|^2)\:k^2 \sin(k\eta-z_k)\nn
G(k\eta,z_k)&=&-\frac{k(A_k^2-|B_k|^2)}{z_k}\nn
&\times&\Big[z_k\cos(k\eta-z_k)+\sin(k\eta-z_k) \Big]\nn
\earr
On the other hand,
\barr\label{modykgk}
|y_k(\tc)|^2&=& \frac{(A_k^2+|B_k|^2)(1+z_k^2)}{z_k^2} +2 |A_k| |B_k|(z_k^2-1)\nn
&\times&\cos(2z_k+\beta)-4 |A_k| |B_k| z_k \sin(2z_k+\beta)\nn
|g_k(\tc)|^2&=&k^2 \Big[A_k^2+|B_k|^2-2|A_k||B_k|\times\nn
&\times&\cos(2z_k+\beta) \Big]
\earr
Remember that $A_k$ and $B_k$ [Eqs. \eqref{Ak} and \eqref{Bk}] have dependence on $z_0\equiv k\eta_0$, the initial condition of the vacuum state chosen.

\section{The $|a_{lm}|_{ML}^2$ and the explicit form of $Q (z_0, z_k)$}\label{appB}

Since we are interested in the observational relevant modes, which have wavelengths greater than the Hubble radius during inflation, as we mentioned in section \ref{colapsos}, we will take the limit $|k\eta|\to0$ in the functions $F$ and $G$ given in Eqs. \eqref{fg}, and we will define $M$ and $N$ as:
\barr\label{MN}
&&M(z_0, z_k)\equiv \lim_{|k\eta|\to 0} F(k\eta, z_k)\:|y_k(\tc)|\nn
&&N(z_0, z_k)\equiv \lim_{|k\eta|\to 0} G(k\eta, z_k) \:|g_k(\tc)|
\earr
Note that we have explicitly written that there is a dependence on $z_0=k\eta_0$ in the functions $M$ and $N$.

The explicit expressions for $|a_{lm}|^2_{\text{ML}}$ can be found by substituting $\mR_\nk$, given in Eq. \eqref{Rkposta}, into Eq. \eqref{alm2} and then making the identification $|a_{lm}|^2_{\text{ML}}=\overline{|a_{lm}|^2}$. This is,
\barr\label{alm2ML1}
|a_{lm}|^{2}_{\text{ML}} &=& \frac{16 \pi^2}{L^6}\sum_{\nk,\nk'}  j_l (kR_D) j_l(k'R_D)\nn
&\times& Y_{lm}^{*}(\hat{k}) Y_{lm} (\hat{k}') T(k) T(k') \overline{\mR_\nk \mR_{\nk'}^{*}}
\earr
Therefore, it is
\barr\label{alm2ML2}
&&|a_{lm}|^{2}_{\text{ML}} = \frac{8 \pi^2 H^2}{L^3 \epsilon_1 M_P^2}\sum_{\nk,\nk'} \frac{ j_l (kR_D) j_l(k'R_D) }{k^2\:k'^2}\nn
&\times& Y_{lm}^{*}(\hat{k}) Y_{lm} (\hat{k}') T(k) T(k')\Big[\lambda_1^2 M(z_0,z_k)M(z_0,z_{k'})\nn
&\times& \overline{X_{\nk,1} X_{\nk',1}^{*}}+\lambda_2^2\: N(z_0,z_k)  N(z_0,z_{k'}) \overline{X_{\nk,2} X_{\nk',2}^{*}}\Big]
\earr
where we have used that the random variables $X_{\nk,1}$ and $X_{\nk,2}$ are uncorrelated, so $\overline{X_{\nk,1} X_{\nk',2}^{*}}=0=\overline{X_{\nk,2} X_{\nk',1}^{*}}$.

From Eq. \eqref{alm2ML2}, and since we are assuming that the $x_{\nk}^{R,I}$ variables are uncorrelated, and thus the ensemble average of the product of these random variables satisfies Eq. \eqref{distribucionx}, we can finally identify the function $Q (z_0, z_k)$ with
\beq
Q(z_0, z_k)\equiv\frac{1}{k}\Big[\lambda_1^2\:M^2(z_0, z_k)+ \lambda_2^2\:N^2(z_0, z_k)\Big]\nn
\eeq
which explicitly turns out to be:
\barr\label{Qposta}
&&Q(z_0, z_k)=\frac{1}{16\:z_k^2\: z_0^2}\Bigg\{\lambda_2^2\:\Big[1+2z_0^2-z_0\sqrt{4+\frac{1}{z_0^2}}\nn
&\times&\cos[2z_k-2z_0+\arctan(2z_0)]\Big] [z_k \cos(-z_k)+\sin(-z_k)]^2\nn
&+&\lambda_1^2\:\sin^2(-z_k)\Bigg[(1+z_k^2)(1+2z_0^2)+z_0 \sqrt{4+\frac{1}{z_0^2}}\nn
&\times&\Big[(z_k^2-1)\cos[2z_k-2z_0+\arctan(2z_0)]\nn
&-&2z_k\sin[2z_k-2z_0+\arctan(2z_0)]\Big]\Bigg]\Bigg\}\nn
\earr

\section{Equivalent power spectra $P(k)$}\label{appC}

In the traditional inflationary scenario, the power spectrum $P(k)$ is obtained by computing the quantum two-point correlation function. That is, if $\hat{\mR}$ represents the quantum field associated to the scalar metric perturbation, then the power spectrum is taken to be
\beq
\bra 0 | \hat \mR_{\nk} \hat \mR_{\nk'}^{*} | 0 \ket = \frac{2 \pi^2}{k^3} P(k)
\delta(\nk-\nk')
\eeq
On the other hand, let us recall that in general, the definition of the power spectrum is given in terms of $\mR_{\nk}$, i.e. a classical stochastic field and not a quantum field. Therefore, the standard approach is based on the identification:
\beq\label{relacion}
\bra 0 | \hat \mR_{\nk} \hat \mR_{\nk'}^{*} | 0 \ket = \overline{\mR_{\nk} \mR_{\nk'}^{*}}
\eeq
with $\overline{\mR_{\nk}\mR_{\nk'}^{*}}$ denoting an average over an ensemble of classical stochastic fields. The justification for the relation above relies on arguments based on decoherence and the squeezing nature of the evolved vacuum state \cite{grishchuk, kiefer} (although we do not subscribe to such arguments for the reasons exposed in \cite{LLS13,Shortcomings}). It is also important to mention that based on the above hypotheses, there is a strong matching between the predictions based on the standard approach and the observational data.

On the other hand, in our proposal, the procedure to obtain an equivalent power spectrum is different from the traditional approach. We start by focusing on the temperature
anisotropies of the CMB observed today on the celestial two-sphere and its relation to the scalar metric perturbation $\mR$. In Fourier space, this relation can be written as (see Sect. \ref{espectro}),
\begin{equation}\label{deltaT}
\Theta (\nk) = T(k) \mR_{\nk}
\end{equation}
where $T(k)$ is known as the transfer function, which contains the physics between the beginning of the radiation-dominated era and the present, i.e the modifications associated with late-time physics. [A well known result (the \emph{Sachs-Wolfe effect}) is, for instance, that $T(k)\simeq1/5$ for very large scales].

On the other hand, the observational data are described in terms of the coefficients $a_{lm}$ of the multipolar series expansion
\begin{equation}\label{expansion.alpha}
\begin{split}
\frac{\delta T}{T_0}(\theta,\varphi)=\sum_{lm}a_{lm}Y_{lm}(\theta,\varphi),
\\
a_{lm}=\int
\frac{\delta T}{T_0}(\theta,\varphi)Y^*_{lm}(\theta,\varphi)d\Omega,
\end{split}
\end{equation}
here $\theta$ and $\varphi$ are the coordinates on the celestial two-sphere, with $Y_{lm}(\theta,\varphi)$ as the spherical harmonics.

The values for the quantities $a_{lm}$ are then given by
\begin{equation}\label{alm2p}
a_{lm} = \frac{4 \pi i^l}{3}   \int \frac{d^3{k}}{(2 \pi)^3} j_l (kR_D)
Y_{lm}^* (\hat{k}) T (k) \mR_{\nk}
\end{equation}
with $j_l (kR_D)$ being the spherical Bessel function of order $l$, and $R_D$ is the comoving radius of the last scattering surface. The metric perturbation $\mR_{\nk}$ is
the primordial curvature perturbation (in the comoving gauge).

By using Eq. \eqref{masterpi} (with $\mR_{\nk}\simeq\Psi_{\nk}/\epsilon_1$) into Eq. \eqref{alm2p} we obtain
\beq\label{almd1}
a_{lm}= \frac{4 \pi i^l }{3}  \frac{H}{\sqrt{2 \epsilon_1}M_P} \int
\frac{d^3{k}}{(2 \pi)^3} j_l (kR_D)
Y_{lm}^* (\hat{k}) T (k) \frac{\bra \hat \pi_{\nk} \ket}{k^2}.
\eeq
The previous expression shows how the expectation value of the momentum field in the post-collapse state acts as a source for the coefficients $a_{lm}$.

Furthermore, the angular power spectrum is defined by
\beq
C_{l} = \frac{1}{2l+1} \sum_m |a_{lm}|^2.
\eeq
For the reasons presented in Sect. \ref{espectro}, we can identify the observed value $|a_{lm}|^2$ with the most likely value of $|a_{lm}|^2_{ML}$ and in turn, assume that the most likely value coincides approximately with the average $\overline{|a_{lm}|^2}$.

Thus, in our approach, the observed $C_l$ coincides with
\beq
C_{l} \simeq  \frac{1}{2l+1} \sum_m \overline{|a_{lm}|^2}.
\eeq
From Eq. \eqref{almd1} we obtain
\barr\label{almd2}
 \overline{|a_{lm}|^2} &=& \left (\frac{4 \pi}{3} \right)^2   \int
 \frac{d^3{k} d^3{k'}}{(2 \pi)^6} j_l (kR_D) j_l (k'R_D)
\nn
 &\times&  Y_{lm}^* (\hat{k}) Y_{lm} (\hat k')  T (k) T(k')
\overline{\mR_{\nk} \mR_{\nk'}^{*}} \nn
 &=& \left (\frac{4 \pi}{3} \right)^2   \int
 \frac{d^3{k} d^3{k'}}{(2 \pi)^6} j_l (kR_D) j_l (k'R_D)
 \nn
 &\times&  Y_{lm}^* (\hat{k}) Y_{lm} (\hat k') T(k) T(k') \bigg[
\frac{H^2}{2 \epsilon_1M_P^2}
 \overline{\frac{\bra \hat \pi_{\nk} \ket \bra \hat \pi_{\nk'} \ket^{*} }{k^2
k^{'2}}} \bigg]. \nn
\earr

Consequently using the generic definition of the power spectrum,
\beq
\overline{\mR_{\nk} \mR_{\nk'}^{*}} \equiv \frac{2 \pi^2}{k^3} P(k)
\delta(\nk-\nk')
\eeq
and also using Eq. \eqref{almd2}, the power spectrum, associated to $\mR_{\nk}$, in our approach is given by
\beq
P(k) = \frac{H^2}{4 \pi^2\epsilon_1 M_P^2}
\frac{ \overline{\bra \hat \pi_{\nk} (\eta) \ket \bra \hat \pi_{\nk} (\eta) \ket^{*}}}{k}.
\eeq
The quantity $\overline{\bra \hat \pi_{\nk} (\eta) \ket \bra\hat \pi_{\nk} (\eta) \ket^{*} }$ is obtained by using Eq. \eqref{expecpitexto} in the limit $-k\eta \to 0$, i.e. when the proper wavelength of the modes of interest are bigger than the Hubble radius.

\section{A non-scale invariant power spectrum for $|z_0|\to 0$}\label{appD}

In this Appendix, we will show that if $|z_0|\to 0$, in both approaches, i.e. in the standard inflationary model and in our picture with the additional collapse hypothesis, then the resulting shape of the scalar power spectrum $\mathcal{P}(k)$ is not consistent with the observational data.

In the standard approach, the modes $v_k$ of the Mukhanov-Sasaki variable satisfy \cite{Mukhanov1992, Mukhanov2005},
\beq\label{vkmuk}
v''_k+\Big(k^2-\frac{z''}{z}\Big)v_k=0
\eeq

Note that the equation of motion for $y_k(\eta)$, Eq. \eqref{ykmov2}, is identical to Eq. \eqref{vkmuk}. That is, when neglecting the slow roll parameters $\frac{z''}{z}\simeq\frac{a''}{a} =\frac{2}{\eta^2}$. As we know, a general solution in such a case will be,
\beq\label{vkpostaApp}
v_k(\eta)=A_k \Big(1-\frac{1}{k\eta}\Big) e^{-i k\eta}+B_k\Big(1+\frac{1}{k\eta}\Big) e^{i k\eta}
\eeq
where we assumed $A_k\in \mathbb{R}$ and $B_k\in \mathbb{C}$.

On the other hand, in the standard scenario, the (dimensional) scalar power spectrum is obtained from
\beq\label{Pmuk}
\mathcal{P}(k)\simeq \lim_{-k\eta\to 0}\frac{|v_k(\eta)|^2}{M_P^2 \epsilon_1 a^2(\eta)}
\eeq
By using Eq. \eqref{vkpostaApp} into Eq. \eqref{Pmuk} one obtains
\beq\label{Pkmuk2}
\mathcal{P}(k)\simeq \frac{H^2}{M_P^2 \epsilon_1 k^2}\Big[A_k^2+|B_k|^2-2A_k\textrm{Re}(B_k)\Big]
\eeq

From Eqs. \eqref{Ak} and \eqref{Bk} we have seen that if $z_0\to -\infty$, the initial conditions provided by the BD vacuum are recovered, which implies that, $A_k=\frac{1}{\sqrt{2k}}$ and $B_k=0$. In that case, Eq. \eqref{Pkmuk2} implies $\mathcal{P}(k)\propto\frac{1}{k^3}$, so a scale invariant power spectrum is obtained.

However, if $|z_0|\to 0$ then we can make the following approximations in Eqs. \eqref{Ak} and \eqref{Bk}:
\barr\label{ABappox}
&&|A_k|^2\simeq\frac{1}{8k|z_0|^2} \quad |B_k|^2=\frac{1}{8k|z_0|^2}\\
&&A_k\simeq\frac{1}{\sqrt{8k}|z_0|} \quad  \textrm{Re}(B_k)\simeq-\frac{1}{\sqrt{8k}|z_0|}
\earr
and since $z_0\equiv k\eta_0$, we obtain from Eq. \eqref{Pkmuk2},
\beq\label{Pkmukaprox}
\mathcal{P}(k)\simeq\frac{H^2}{8k^3}\frac{1}{|z_0|^2}\propto\frac{1}{k^5}
\eeq
Therefore, if $|z_0|\to 0$, then the shape of the power spectrum, which resulted from choosing a vacuum state such that it minimizes the renormalized stress-energy tensor via a Hadamard procedure, is not compatible with the observational data. The initial conditions obtained from the vacuum choice fix the value of $A_k$ and $B_k$ given by Eqs. \eqref{Ak} and \eqref{Bk}.

Now, let us show that under our approach, with the collapse hypothesis included, we arrive at the same result. For simplicity, we will assume $\lambda_1=1=\lambda_2$.

For $|z_0|\to 0$, and taking into account that $|z_k|<|z_0|$, we perform a Taylor expansion in Eq. \eqref{Qposta}. At the leading order in $|z_0|$ and $|z_k|$, we obtain:
\beq\label{Qaprox}
Q(z_0, z_k)\simeq\frac{1}{8|z_0|^2}+\frac{|z_k|^2}{12|z_0|^2}+\mathcal{O}(|z_0|,|z_k|^4)
\eeq
Again, keeping the first relevant term in Eq. \eqref{Qaprox}, and since $z_0=k\eta_0$ we finally arrive at
\beq\label{Pkourprox}
\mathcal{P}(k)\propto\frac{1}{k^5}
\eeq
We observe that the result is independent of the parametrization for $\tc$. Therefore, as in the standard case, the shape of the power spectrum is not consistent with the data. We attained this negative result not due to the collapse hypothesis but because of the initial conditions, provided by the novel choice of the vacuum state, when $|z_0| \to 0$.

\bibliography{bibliografia}

\begin{thebibliography}{69}
\expandafter\ifx\csname natexlab\endcsname\relax\def\natexlab#1{#1}\fi
\expandafter\ifx\csname bibnamefont\endcsname\relax
  \def\bibnamefont#1{#1}\fi
\expandafter\ifx\csname bibfnamefont\endcsname\relax
  \def\bibfnamefont#1{#1}\fi
\expandafter\ifx\csname citenamefont\endcsname\relax
  \def\citenamefont#1{#1}\fi
\expandafter\ifx\csname url\endcsname\relax
  \def\url#1{\texttt{#1}}\fi
\expandafter\ifx\csname urlprefix\endcsname\relax\def\urlprefix{URL }\fi
\providecommand{\bibinfo}[2]{#2}
\providecommand{\eprint}[2][]{\url{#2}}

\bibitem[{\citenamefont{Starobinsky}(1980)}]{starobinsky}
\bibinfo{author}{\bibfnamefont{A.~A.} \bibnamefont{Starobinsky}},
  \bibinfo{journal}{Phys. Lett.} \textbf{\bibinfo{volume}{B91}},
  \bibinfo{pages}{99} (\bibinfo{year}{1980}).

\bibitem[{\citenamefont{Guth}(1981)}]{guth}
\bibinfo{author}{\bibfnamefont{A.~H.} \bibnamefont{Guth}},
  \bibinfo{journal}{Phys. Rev.} \textbf{\bibinfo{volume}{D23}},
  \bibinfo{pages}{347} (\bibinfo{year}{1981}).

\bibitem[{\citenamefont{Linde}(1982)}]{linde}
\bibinfo{author}{\bibfnamefont{A.~D.} \bibnamefont{Linde}},
  \bibinfo{journal}{Phys. Lett.} \textbf{\bibinfo{volume}{B108}},
  \bibinfo{pages}{389} (\bibinfo{year}{1982}).

\bibitem[{\citenamefont{Albrecht and Steinhardt}(1982)}]{albrecht}
\bibinfo{author}{\bibfnamefont{A.}~\bibnamefont{Albrecht}} \bibnamefont{and}
  \bibinfo{author}{\bibfnamefont{P.~J.} \bibnamefont{Steinhardt}},
  \bibinfo{journal}{Phys. Rev. Lett.} \textbf{\bibinfo{volume}{48}},
  \bibinfo{pages}{1220} (\bibinfo{year}{1982}).

\bibitem[{\citenamefont{Mukhanov and Chibisov}(1981)}]{mukhanov81}
\bibinfo{author}{\bibfnamefont{V.~F.} \bibnamefont{Mukhanov}} \bibnamefont{and}
  \bibinfo{author}{\bibfnamefont{G.~V.} \bibnamefont{Chibisov}},
  \bibinfo{journal}{JETP Lett.} \textbf{\bibinfo{volume}{33}},
  \bibinfo{pages}{532} (\bibinfo{year}{1981}).

\bibitem[{\citenamefont{Mukhanov and Chibisov}(1982)}]{mukhanov2}
\bibinfo{author}{\bibfnamefont{V.~F.} \bibnamefont{Mukhanov}} \bibnamefont{and}
  \bibinfo{author}{\bibfnamefont{G.}~\bibnamefont{Chibisov}},
  \bibinfo{journal}{Sov. Phys. JETP} \textbf{\bibinfo{volume}{56}},
  \bibinfo{pages}{258} (\bibinfo{year}{1982}).

\bibitem[{\citenamefont{Starobinsky}(1983)}]{starobinsky2}
\bibinfo{author}{\bibfnamefont{A.~A.} \bibnamefont{Starobinsky}},
  \bibinfo{journal}{Sov.Astron.Lett.} \textbf{\bibinfo{volume}{9}},
  \bibinfo{pages}{302} (\bibinfo{year}{1983}).

\bibitem[{\citenamefont{Hawking and Moss}(1983)}]{hawking}
\bibinfo{author}{\bibfnamefont{S.~W.} \bibnamefont{Hawking}} \bibnamefont{and}
  \bibinfo{author}{\bibfnamefont{I.~G.} \bibnamefont{Moss}},
  \bibinfo{journal}{Nucl. Phys.} \textbf{\bibinfo{volume}{B224}},
  \bibinfo{pages}{180} (\bibinfo{year}{1983}).

\bibitem[{\citenamefont{Hawking}(1982)}]{hawking2}
\bibinfo{author}{\bibfnamefont{S.~W.} \bibnamefont{Hawking}},
  \bibinfo{journal}{Phys. Lett.} \textbf{\bibinfo{volume}{B115}},
  \bibinfo{pages}{295} (\bibinfo{year}{1982}).

\bibitem[{\citenamefont{Ade et~al.}(2016{\natexlab{a}})}]{planck2015}
\bibinfo{author}{\bibfnamefont{P.~A.~R.} \bibnamefont{Ade}}
  \bibnamefont{et~al.} (\bibinfo{collaboration}{Planck}),
  \bibinfo{journal}{Astron. Astrophys.} \textbf{\bibinfo{volume}{594}},
  \bibinfo{pages}{A13} (\bibinfo{year}{2016}{\natexlab{a}}).

\bibitem[{\citenamefont{Aghanim et~al.}(2016)}]{planck2015likelihoods}
\bibinfo{author}{\bibfnamefont{N.}~\bibnamefont{Aghanim}} \bibnamefont{et~al.}
  (\bibinfo{collaboration}{Planck}), \bibinfo{journal}{Astron. Astrophys.}
  \textbf{\bibinfo{volume}{594}}, \bibinfo{pages}{A11} (\bibinfo{year}{2016}).

\bibitem[{\citenamefont{Ade et~al.}(2016{\natexlab{b}})}]{plkinflation15}
\bibinfo{author}{\bibfnamefont{P.~A.~R.} \bibnamefont{Ade}}
  \bibnamefont{et~al.} (\bibinfo{collaboration}{Planck}),
  \bibinfo{journal}{Astron. Astrophys.} \textbf{\bibinfo{volume}{594}},
  \bibinfo{pages}{A20} (\bibinfo{year}{2016}{\natexlab{b}}),
  \eprint{1502.02114}.

\bibitem[{\citenamefont{Martin et~al.}(2014)\citenamefont{Martin, Ringeval,
  Trotta, and Vennin}}]{Martin14}
\bibinfo{author}{\bibfnamefont{J.}~\bibnamefont{Martin}},
  \bibinfo{author}{\bibfnamefont{C.}~\bibnamefont{Ringeval}},
  \bibinfo{author}{\bibfnamefont{R.}~\bibnamefont{Trotta}}, \bibnamefont{and}
  \bibinfo{author}{\bibfnamefont{V.}~\bibnamefont{Vennin}},
  \bibinfo{journal}{JCAP} \textbf{\bibinfo{volume}{1403}}, \bibinfo{pages}{039}
  (\bibinfo{year}{2014}).

\bibitem[{\citenamefont{{Perez} et~al.}(2005)\citenamefont{{Perez}, {Sahlmann},
  and {Sudarsky}}}]{PSS06}
\bibinfo{author}{\bibfnamefont{A.}~\bibnamefont{{Perez}}},
  \bibinfo{author}{\bibfnamefont{H.}~\bibnamefont{{Sahlmann}}},
  \bibnamefont{and}
  \bibinfo{author}{\bibfnamefont{D.}~\bibnamefont{{Sudarsky}}},
  \bibinfo{journal}{Classical and Quantum Gravity}
  \textbf{\bibinfo{volume}{23}}, \bibinfo{pages}{2317} (\bibinfo{year}{2005}),
  \eprint{gr-qc/0508100}.

\bibitem[{\citenamefont{{Sudarsky}}(2011)}]{Shortcomings}
\bibinfo{author}{\bibfnamefont{D.}~\bibnamefont{{Sudarsky}}},
  \bibinfo{journal}{International Journal of Modern Physics D}
  \textbf{\bibinfo{volume}{20}}, \bibinfo{pages}{509} (\bibinfo{year}{2011}),
  \eprint{0906.0315}.

\bibitem[{\citenamefont{{Landau} et~al.}(2013)\citenamefont{{Landau},
  {Le{\'o}n}, and {Sudarsky}}}]{LLS13}
\bibinfo{author}{\bibfnamefont{S.}~\bibnamefont{{Landau}}},
  \bibinfo{author}{\bibfnamefont{G.}~\bibnamefont{{Le{\'o}n}}},
  \bibnamefont{and}
  \bibinfo{author}{\bibfnamefont{D.}~\bibnamefont{{Sudarsky}}},
  \bibinfo{journal}{Physical Review D} \textbf{\bibinfo{volume}{88}},
  \bibinfo{eid}{023526} (\bibinfo{year}{2013}), \eprint{1107.3054}.

\bibitem[{\citenamefont{{Polarski} and {Starobinsky}}(1996)}]{polarski}
\bibinfo{author}{\bibfnamefont{D.}~\bibnamefont{{Polarski}}} \bibnamefont{and}
  \bibinfo{author}{\bibfnamefont{A.~A.} \bibnamefont{{Starobinsky}}},
  \bibinfo{journal}{Classical and Quantum Gravity}
  \textbf{\bibinfo{volume}{13}}, \bibinfo{pages}{377} (\bibinfo{year}{1996}),
  \eprint{gr-qc/9504030}.

\bibitem[{\citenamefont{Kiefer and Polarski}(2009)}]{kiefer}
\bibinfo{author}{\bibfnamefont{C.}~\bibnamefont{Kiefer}} \bibnamefont{and}
  \bibinfo{author}{\bibfnamefont{D.}~\bibnamefont{Polarski}},
  \bibinfo{journal}{Adv. Sci. Lett.} \textbf{\bibinfo{volume}{2}},
  \bibinfo{pages}{164} (\bibinfo{year}{2009}), \eprint{0810.0087}.

\bibitem[{\citenamefont{{de Un{\'a}nue} and {Sudarsky}}(2008)}]{US08}
\bibinfo{author}{\bibfnamefont{A.}~\bibnamefont{{de Un{\'a}nue}}}
  \bibnamefont{and}
  \bibinfo{author}{\bibfnamefont{D.}~\bibnamefont{{Sudarsky}}},
  \bibinfo{journal}{Physical Review D} \textbf{\bibinfo{volume}{78}},
  \bibinfo{pages}{043510} (\bibinfo{year}{2008}), \eprint{arXiv:0801.4702}.

\bibitem[{\citenamefont{{Diez-Tejedor} and {Sudarsky}}(2012)}]{DT11}
\bibinfo{author}{\bibfnamefont{A.}~\bibnamefont{{Diez-Tejedor}}}
  \bibnamefont{and}
  \bibinfo{author}{\bibfnamefont{D.}~\bibnamefont{{Sudarsky}}},
  \bibinfo{journal}{JCAP} \textbf{\bibinfo{volume}{7}}, \bibinfo{eid}{045}
  (\bibinfo{year}{2012}), \eprint{1108.4928}.

\bibitem[{\citenamefont{Le\'{o}n et~al.}(2014)\citenamefont{Le\'{o}n, Landau,
  and Piccirilli}}]{pia}
\bibinfo{author}{\bibfnamefont{G.}~\bibnamefont{Le\'{o}n}},
  \bibinfo{author}{\bibfnamefont{S.~J.} \bibnamefont{Landau}},
  \bibnamefont{and} \bibinfo{author}{\bibfnamefont{M.~P.}
  \bibnamefont{Piccirilli}}, \bibinfo{journal}{Phys. Rev.}
  \textbf{\bibinfo{volume}{D 90}}, \bibinfo{pages}{083525}
  (\bibinfo{year}{2014}), \eprint{1410.1562}.

\bibitem[{\citenamefont{{Ca{\~n}ate} et~al.}(2013)\citenamefont{{Ca{\~n}ate},
  {Pearle}, and {Sudarsky}}}]{CPS13}
\bibinfo{author}{\bibfnamefont{P.}~\bibnamefont{{Ca{\~n}ate}}},
  \bibinfo{author}{\bibfnamefont{P.}~\bibnamefont{{Pearle}}}, \bibnamefont{and}
  \bibinfo{author}{\bibfnamefont{D.}~\bibnamefont{{Sudarsky}}},
  \bibinfo{journal}{Phys.Rev. D} \textbf{\bibinfo{volume}{87}},
  \bibinfo{eid}{104024} (\bibinfo{year}{2013}), \eprint{1211.3463}.

\bibitem[{\citenamefont{Martin et~al.}(2012)\citenamefont{Martin, Vennin, and
  Peter}}]{jmartin}
\bibinfo{author}{\bibfnamefont{J.}~\bibnamefont{Martin}},
  \bibinfo{author}{\bibfnamefont{V.}~\bibnamefont{Vennin}}, \bibnamefont{and}
  \bibinfo{author}{\bibfnamefont{P.}~\bibnamefont{Peter}},
  \bibinfo{journal}{Phys.Rev.} \textbf{\bibinfo{volume}{D86}},
  \bibinfo{pages}{103524} (\bibinfo{year}{2012}), \eprint{1207.2086}.

\bibitem[{\citenamefont{Das et~al.}(2013)\citenamefont{Das, Lochan, Sahu, and
  Singh}}]{tpsingh}
\bibinfo{author}{\bibfnamefont{S.}~\bibnamefont{Das}},
  \bibinfo{author}{\bibfnamefont{K.}~\bibnamefont{Lochan}},
  \bibinfo{author}{\bibfnamefont{S.}~\bibnamefont{Sahu}}, \bibnamefont{and}
  \bibinfo{author}{\bibfnamefont{T.}~\bibnamefont{Singh}},
  \bibinfo{journal}{Phys.Rev.} \textbf{\bibinfo{volume}{D88}},
  \bibinfo{pages}{085020} (\bibinfo{year}{2013}), \eprint{1304.5094}.

\bibitem[{\citenamefont{Leon and Bengochea}(2016)}]{LB2015}
\bibinfo{author}{\bibfnamefont{G.}~\bibnamefont{Leon}} \bibnamefont{and}
  \bibinfo{author}{\bibfnamefont{G.~R.} \bibnamefont{Bengochea}},
  \bibinfo{journal}{Eur. Phys. J.} \textbf{\bibinfo{volume}{C76}},
  \bibinfo{pages}{29} (\bibinfo{year}{2016}), \eprint{1502.04907}.

\bibitem[{\citenamefont{Alexander et~al.}(2016)\citenamefont{Alexander, Jyoti,
  and Magueijo}}]{magueijo2016}
\bibinfo{author}{\bibfnamefont{S.}~\bibnamefont{Alexander}},
  \bibinfo{author}{\bibfnamefont{D.}~\bibnamefont{Jyoti}}, \bibnamefont{and}
  \bibinfo{author}{\bibfnamefont{J.}~\bibnamefont{Magueijo}},
  \bibinfo{journal}{Phys. Rev.} \textbf{\bibinfo{volume}{D94}},
  \bibinfo{pages}{043502} (\bibinfo{year}{2016}), \eprint{1602.01216}.

\bibitem[{\citenamefont{Le\'{o}n et~al.}(2015)\citenamefont{Le\'{o}n,
  Kraiselburd, and Landau}}]{lucila}
\bibinfo{author}{\bibfnamefont{G.}~\bibnamefont{Le\'{o}n}},
  \bibinfo{author}{\bibfnamefont{L.}~\bibnamefont{Kraiselburd}},
  \bibnamefont{and} \bibinfo{author}{\bibfnamefont{S.~J.}
  \bibnamefont{Landau}}, \bibinfo{journal}{Phys. Rev.}
  \textbf{\bibinfo{volume}{D92}}, \bibinfo{pages}{083516}
  (\bibinfo{year}{2015}), \eprint{1509.08399}.

\bibitem[{\citenamefont{Le\'on et~al.}(2016)\citenamefont{Le\'on, Majhi, Okon,
  and Sudarsky}}]{abhishek}
\bibinfo{author}{\bibfnamefont{G.}~\bibnamefont{Le\'on}},
  \bibinfo{author}{\bibfnamefont{A.}~\bibnamefont{Majhi}},
  \bibinfo{author}{\bibfnamefont{E.}~\bibnamefont{Okon}}, \bibnamefont{and}
  \bibinfo{author}{\bibfnamefont{D.}~\bibnamefont{Sudarsky}},
  \emph{\bibinfo{title}{{Reassessing the link between B-modes and inflation}}}
  (\bibinfo{year}{2016}), \eprint{1607.03523}.

\bibitem[{\citenamefont{Benetti et~al.}(2016)\citenamefont{Benetti, Landau, and
  Alcaniz}}]{benetti06}
\bibinfo{author}{\bibfnamefont{M.}~\bibnamefont{Benetti}},
  \bibinfo{author}{\bibfnamefont{S.~J.} \bibnamefont{Landau}},
  \bibnamefont{and} \bibinfo{author}{\bibfnamefont{J.~S.}
  \bibnamefont{Alcaniz}}, \bibinfo{journal}{JCAP}
  \textbf{\bibinfo{volume}{1612}}, \bibinfo{pages}{035} (\bibinfo{year}{2016}),
  \eprint{1610.03091}.

\bibitem[{\citenamefont{Hollands and Wald}(2015)}]{WaldQFTCS}
\bibinfo{author}{\bibfnamefont{S.}~\bibnamefont{Hollands}} \bibnamefont{and}
  \bibinfo{author}{\bibfnamefont{R.~M.} \bibnamefont{Wald}},
  \bibinfo{journal}{Phys. Rept.} \textbf{\bibinfo{volume}{574}},
  \bibinfo{pages}{1} (\bibinfo{year}{2015}), \eprint{1401.2026}.

\bibitem[{\citenamefont{Wheeler}(1964)}]{Em1}
\bibinfo{author}{\bibfnamefont{J.~A.} \bibnamefont{Wheeler}}, in
  \emph{\bibinfo{booktitle}{Relativity, Groups, and Topology}}, edited by
  \bibinfo{editor}{\bibfnamefont{B.~S.} \bibnamefont{DeWitt}} \bibnamefont{and}
  \bibinfo{editor}{\bibfnamefont{C.}~\bibnamefont{DeWitt}}
  (\bibinfo{publisher}{Gordon and Breach}, \bibinfo{year}{1964}).

\bibitem[{\citenamefont{Finkelstein}(1969)}]{Em2}
\bibinfo{author}{\bibfnamefont{D.}~\bibnamefont{Finkelstein}},
  \bibinfo{journal}{Phys. Rev.} \textbf{\bibinfo{volume}{184}},
  \bibinfo{pages}{1261} (\bibinfo{year}{1969}).

\bibitem[{\citenamefont{Konopka et~al.}(2008)\citenamefont{Konopka,
  Markopoulou, and Severini}}]{Em3}
\bibinfo{author}{\bibfnamefont{T.}~\bibnamefont{Konopka}},
  \bibinfo{author}{\bibfnamefont{F.}~\bibnamefont{Markopoulou}},
  \bibnamefont{and} \bibinfo{author}{\bibfnamefont{S.}~\bibnamefont{Severini}},
  \bibinfo{journal}{Phys. Rev. D} \textbf{\bibinfo{volume}{77}},
  \bibinfo{pages}{104029} (\bibinfo{year}{2008}).

\bibitem[{\citenamefont{Oriti}(2009)}]{Em4}
\bibinfo{author}{\bibfnamefont{D.}~\bibnamefont{Oriti}}, in
  \emph{\bibinfo{booktitle}{Approaches to Quantum Gravity}}, edited by
  \bibinfo{editor}{\bibfnamefont{D.}~\bibnamefont{Oriti}}
  (\bibinfo{publisher}{Cambridge University Press}, \bibinfo{year}{2009}).

\bibitem[{\citenamefont{Steinacker}(2010)}]{Em5}
\bibinfo{author}{\bibfnamefont{H.}~\bibnamefont{Steinacker}},
  \bibinfo{journal}{Class. Quant Grav.} \textbf{\bibinfo{volume}{27}},
  \bibinfo{pages}{133001} (\bibinfo{year}{2010}).

\bibitem[{\citenamefont{Fulling}(1989)}]{fullingbook}
\bibinfo{author}{\bibfnamefont{S.}~\bibnamefont{Fulling}},
  \emph{\bibinfo{title}{Aspects of Quantum Field Theory in Curved Spacetime}},
  London Mathematical Society St (\bibinfo{publisher}{Cambridge University
  Press}, \bibinfo{year}{1989}), ISBN \bibinfo{isbn}{9780521377683}.

\bibitem[{\citenamefont{Mukhanov et~al.}(1992)\citenamefont{Mukhanov, Feldman,
  and Brandenberger}}]{Mukhanov1992}
\bibinfo{author}{\bibfnamefont{V.~F.} \bibnamefont{Mukhanov}},
  \bibinfo{author}{\bibfnamefont{H.~A.} \bibnamefont{Feldman}},
  \bibnamefont{and} \bibinfo{author}{\bibfnamefont{R.~H.}
  \bibnamefont{Brandenberger}}, \bibinfo{journal}{Phys. Rept.}
  \textbf{\bibinfo{volume}{215}}, \bibinfo{pages}{203} (\bibinfo{year}{1992}).

\bibitem[{\citenamefont{Mukhanov and Winitzki}(2007)}]{MukhanovQFTCS}
\bibinfo{author}{\bibfnamefont{V.}~\bibnamefont{Mukhanov}} \bibnamefont{and}
  \bibinfo{author}{\bibfnamefont{S.}~\bibnamefont{Winitzki}},
  \emph{\bibinfo{title}{{Introduction to Quantum Effects in Gravity}}}
  (\bibinfo{publisher}{{Cambridge University Press}}, \bibinfo{year}{2007}),
  ISBN \bibinfo{isbn}{0521868343}.

\bibitem[{\citenamefont{Fulling}(1979)}]{fulling79}
\bibinfo{author}{\bibfnamefont{S.~A.} \bibnamefont{Fulling}},
  \bibinfo{journal}{Gen. Rel. Grav.} \textbf{\bibinfo{volume}{10}},
  \bibinfo{pages}{807} (\bibinfo{year}{1979}).

\bibitem[{\citenamefont{Armendariz-Picon}(2007)}]{picon2007}
\bibinfo{author}{\bibfnamefont{C.}~\bibnamefont{Armendariz-Picon}},
  \bibinfo{journal}{JCAP} \textbf{\bibinfo{volume}{0702}}, \bibinfo{pages}{031}
  (\bibinfo{year}{2007}), \eprint{astro-ph/0612288}.

\bibitem[{\citenamefont{Handley et~al.}(2016)\citenamefont{Handley, Lasenby,
  and Hobson}}]{handley16}
\bibinfo{author}{\bibfnamefont{W.}~\bibnamefont{Handley}},
  \bibinfo{author}{\bibfnamefont{A.}~\bibnamefont{Lasenby}}, \bibnamefont{and}
  \bibinfo{author}{\bibfnamefont{M.}~\bibnamefont{Hobson}},
  \bibinfo{journal}{Phys. Rev.} \textbf{\bibinfo{volume}{D94}},
  \bibinfo{pages}{024041} (\bibinfo{year}{2016}), \eprint{1607.04148}.

\bibitem[{\citenamefont{Martin and Brandenberger}(2001)}]{Martin2000}
\bibinfo{author}{\bibfnamefont{J.}~\bibnamefont{Martin}} \bibnamefont{and}
  \bibinfo{author}{\bibfnamefont{R.~H.} \bibnamefont{Brandenberger}},
  \bibinfo{journal}{Phys. Rev.} \textbf{\bibinfo{volume}{D63}},
  \bibinfo{pages}{123501} (\bibinfo{year}{2001}), \eprint{hep-th/0005209}.

\bibitem[{\citenamefont{Danielsson}(2002{\natexlab{a}})}]{Danielsson2002}
\bibinfo{author}{\bibfnamefont{U.~H.} \bibnamefont{Danielsson}},
  \bibinfo{journal}{Phys. Rev.} \textbf{\bibinfo{volume}{D66}},
  \bibinfo{pages}{023511} (\bibinfo{year}{2002}{\natexlab{a}}),
  \eprint{hep-th/0203198}.

\bibitem[{\citenamefont{Martin and Ringeval}(2004)}]{Martin2003}
\bibinfo{author}{\bibfnamefont{J.}~\bibnamefont{Martin}} \bibnamefont{and}
  \bibinfo{author}{\bibfnamefont{C.}~\bibnamefont{Ringeval}},
  \bibinfo{journal}{Phys. Rev.} \textbf{\bibinfo{volume}{D69}},
  \bibinfo{pages}{083515} (\bibinfo{year}{2004}), \eprint{astro-ph/0310382}.

\bibitem[{\citenamefont{Komatsu}(2010)}]{Komatsu2010}
\bibinfo{author}{\bibfnamefont{E.}~\bibnamefont{Komatsu}},
  \bibinfo{journal}{Class. Quant. Grav.} \textbf{\bibinfo{volume}{27}},
  \bibinfo{pages}{124010} (\bibinfo{year}{2010}), \eprint{1003.6097}.

\bibitem[{\citenamefont{Agullo et~al.}(2012)\citenamefont{Agullo,
  Navarro-Salas, and Parker}}]{Agullo2011}
\bibinfo{author}{\bibfnamefont{I.}~\bibnamefont{Agullo}},
  \bibinfo{author}{\bibfnamefont{J.}~\bibnamefont{Navarro-Salas}},
  \bibnamefont{and} \bibinfo{author}{\bibfnamefont{L.}~\bibnamefont{Parker}},
  \bibinfo{journal}{JCAP} \textbf{\bibinfo{volume}{1205}}, \bibinfo{pages}{019}
  (\bibinfo{year}{2012}), \eprint{1112.1581}.

\bibitem[{\citenamefont{{Le{\'o}n} and {Sudarsky}}(2010)}]{Leon10}
\bibinfo{author}{\bibfnamefont{G.}~\bibnamefont{{Le{\'o}n}}} \bibnamefont{and}
  \bibinfo{author}{\bibfnamefont{D.}~\bibnamefont{{Sudarsky}}},
  \bibinfo{journal}{Classical and Quantum Gravity}
  \textbf{\bibinfo{volume}{27}}, \bibinfo{eid}{225017} (\bibinfo{year}{2010}),
  \eprint{1003.5950}.

\bibitem[{\citenamefont{Diez-Tejedor et~al.}(2012)\citenamefont{Diez-Tejedor,
  Leon, and Sudarsky}}]{DLS11}
\bibinfo{author}{\bibfnamefont{A.}~\bibnamefont{Diez-Tejedor}},
  \bibinfo{author}{\bibfnamefont{G.}~\bibnamefont{Leon}}, \bibnamefont{and}
  \bibinfo{author}{\bibfnamefont{D.}~\bibnamefont{Sudarsky}},
  \bibinfo{journal}{Gen.Rel.Grav.} \textbf{\bibinfo{volume}{44}},
  \bibinfo{pages}{2965} (\bibinfo{year}{2012}), \eprint{1106.1176}.

\bibitem[{\citenamefont{Leon et~al.}(2015)\citenamefont{Leon, Landau, and
  Piccirilli}}]{gabrielqdesitter}
\bibinfo{author}{\bibfnamefont{G.}~\bibnamefont{Leon}},
  \bibinfo{author}{\bibfnamefont{S.}~\bibnamefont{Landau}}, \bibnamefont{and}
  \bibinfo{author}{\bibfnamefont{M.~P.} \bibnamefont{Piccirilli}},
  \bibinfo{journal}{Eur. Phys. J.} \textbf{\bibinfo{volume}{C75}},
  \bibinfo{pages}{393} (\bibinfo{year}{2015}), \eprint{1502.00921}.

\bibitem[{\citenamefont{Penrose}(1996)}]{penrose1996}
\bibinfo{author}{\bibfnamefont{R.}~\bibnamefont{Penrose}},
  \bibinfo{journal}{Gen.Rel.Grav.} \textbf{\bibinfo{volume}{28}},
  \bibinfo{pages}{581} (\bibinfo{year}{1996}).

\bibitem[{\citenamefont{Brandenberger et~al.}(1993)\citenamefont{Brandenberger,
  Feldman, and Mukhanov}}]{brandenberger1993}
\bibinfo{author}{\bibfnamefont{R.~H.} \bibnamefont{Brandenberger}},
  \bibinfo{author}{\bibfnamefont{H.}~\bibnamefont{Feldman}}, \bibnamefont{and}
  \bibinfo{author}{\bibfnamefont{V.~F.} \bibnamefont{Mukhanov}}, pp.
  \bibinfo{pages}{19--30} (\bibinfo{year}{1993}), \eprint{astro-ph/9307016}.

\bibitem[{\citenamefont{Armendariz-Picon and Lim}(2003)}]{apicon2003}
\bibinfo{author}{\bibfnamefont{C.}~\bibnamefont{Armendariz-Picon}}
  \bibnamefont{and} \bibinfo{author}{\bibfnamefont{E.~A.} \bibnamefont{Lim}},
  \bibinfo{journal}{JCAP} \textbf{\bibinfo{volume}{0312}}, \bibinfo{pages}{006}
  (\bibinfo{year}{2003}), \eprint{hep-th/0303103}.

\bibitem[{\citenamefont{Danielsson}(2002{\natexlab{b}})}]{dani02}
\bibinfo{author}{\bibfnamefont{U.~H.} \bibnamefont{Danielsson}},
  \bibinfo{journal}{Phys. Rev.} \textbf{\bibinfo{volume}{D66}},
  \bibinfo{pages}{023511} (\bibinfo{year}{2002}{\natexlab{b}}),
  \eprint{hep-th/0203198}.

\bibitem[{\citenamefont{Bohm and Bub}(1966)}]{bohm66}
\bibinfo{author}{\bibfnamefont{D.}~\bibnamefont{Bohm}} \bibnamefont{and}
  \bibinfo{author}{\bibfnamefont{J.}~\bibnamefont{Bub}}, \bibinfo{journal}{Rev.
  Mod. Phys.} \textbf{\bibinfo{volume}{38}}, \bibinfo{pages}{453}
  (\bibinfo{year}{1966}).

\bibitem[{\citenamefont{Pearle}(1976)}]{Pearle76}
\bibinfo{author}{\bibfnamefont{P.~M.} \bibnamefont{Pearle}},
  \bibinfo{journal}{Phys. Rev.} \textbf{\bibinfo{volume}{D13}},
  \bibinfo{pages}{857} (\bibinfo{year}{1976}).

\bibitem[{\citenamefont{Pearle}(1979)}]{Pearle79}
\bibinfo{author}{\bibfnamefont{P.}~\bibnamefont{Pearle}},
  \bibinfo{journal}{International Journal of Theoretical Physics}
  \textbf{\bibinfo{volume}{18}}, \bibinfo{pages}{489} (\bibinfo{year}{1979}).

\bibitem[{\citenamefont{Pearle}(1989)}]{pearle1989}
\bibinfo{author}{\bibfnamefont{P.~M.} \bibnamefont{Pearle}},
  \bibinfo{journal}{Phys.Rev.} \textbf{\bibinfo{volume}{A39}},
  \bibinfo{pages}{2277} (\bibinfo{year}{1989}).

\bibitem[{\citenamefont{Ghirardi et~al.}(1986)\citenamefont{Ghirardi, Rimini,
  and Weber}}]{ghirardi1985}
\bibinfo{author}{\bibfnamefont{G.}~\bibnamefont{Ghirardi}},
  \bibinfo{author}{\bibfnamefont{A.}~\bibnamefont{Rimini}}, \bibnamefont{and}
  \bibinfo{author}{\bibfnamefont{T.}~\bibnamefont{Weber}},
  \bibinfo{journal}{Phys.Rev.} \textbf{\bibinfo{volume}{D34}},
  \bibinfo{pages}{470} (\bibinfo{year}{1986}).

\bibitem[{\citenamefont{Diosi}(1987)}]{diosi1987}
\bibinfo{author}{\bibfnamefont{L.}~\bibnamefont{Diosi}},
  \bibinfo{journal}{Phys.Lett.} \textbf{\bibinfo{volume}{A120}},
  \bibinfo{pages}{377} (\bibinfo{year}{1987}).

\bibitem[{\citenamefont{Diosi}(1989)}]{diosi1989}
\bibinfo{author}{\bibfnamefont{L.}~\bibnamefont{Diosi}},
  \bibinfo{journal}{Phys.Rev.} \textbf{\bibinfo{volume}{A40}},
  \bibinfo{pages}{1165} (\bibinfo{year}{1989}).

\bibitem[{\citenamefont{{Landau} et~al.}(2012)\citenamefont{{Landau},
  {Sc{\'o}ccola}, and {Sudarsky}}}]{LSS12}
\bibinfo{author}{\bibfnamefont{S.~J.} \bibnamefont{{Landau}}},
  \bibinfo{author}{\bibfnamefont{C.~G.} \bibnamefont{{Sc{\'o}ccola}}},
  \bibnamefont{and}
  \bibinfo{author}{\bibfnamefont{D.}~\bibnamefont{{Sudarsky}}},
  \bibinfo{journal}{Physical Review D} \textbf{\bibinfo{volume}{85}},
  \bibinfo{eid}{123001} (\bibinfo{year}{2012}), \eprint{1112.1830}.

\bibitem[{\citenamefont{Das et~al.}(2014)\citenamefont{Das, Sahu, Banerjee, and
  Singh}}]{dasGW}
\bibinfo{author}{\bibfnamefont{S.}~\bibnamefont{Das}},
  \bibinfo{author}{\bibfnamefont{S.}~\bibnamefont{Sahu}},
  \bibinfo{author}{\bibfnamefont{S.}~\bibnamefont{Banerjee}}, \bibnamefont{and}
  \bibinfo{author}{\bibfnamefont{T.}~\bibnamefont{Singh}},
  \bibinfo{journal}{Phys.Rev.} \textbf{\bibinfo{volume}{D90}},
  \bibinfo{pages}{043503} (\bibinfo{year}{2014}), \eprint{1404.5740}.

\bibitem[{\citenamefont{{Le{\'o}n} et~al.}(2011)\citenamefont{{Le{\'o}n}, {De
  Un{\'a}nue}, and {Sudarsky}}}]{Leon11}
\bibinfo{author}{\bibfnamefont{G.}~\bibnamefont{{Le{\'o}n}}},
  \bibinfo{author}{\bibfnamefont{A.}~\bibnamefont{{De Un{\'a}nue}}},
  \bibnamefont{and}
  \bibinfo{author}{\bibfnamefont{D.}~\bibnamefont{{Sudarsky}}},
  \bibinfo{journal}{Classical and Quantum Gravity}
  \textbf{\bibinfo{volume}{28}}, \bibinfo{pages}{155010}
  (\bibinfo{year}{2011}), \eprint{1012.2419}.

\bibitem[{\citenamefont{Deruelle and Mukhanov}(1995)}]{Deruelle1995}
\bibinfo{author}{\bibfnamefont{N.}~\bibnamefont{Deruelle}} \bibnamefont{and}
  \bibinfo{author}{\bibfnamefont{V.~F.} \bibnamefont{Mukhanov}},
  \bibinfo{journal}{Phys.Rev.} \textbf{\bibinfo{volume}{D52}},
  \bibinfo{pages}{5549} (\bibinfo{year}{1995}), \eprint{gr-qc/9503050}.

\bibitem[{\citenamefont{Mukhanov}(2005)}]{Mukhanov2005}
\bibinfo{author}{\bibfnamefont{V.}~\bibnamefont{Mukhanov}},
  \emph{\bibinfo{title}{Physical Foundations of Cosmology}}
  (\bibinfo{publisher}{Cambridge University Press, New York},
  \bibinfo{year}{2005}).

\bibitem[{\citenamefont{Lewis et~al.}(2000)\citenamefont{Lewis, Challinor, and
  Lasenby}}]{CAMB}
\bibinfo{author}{\bibfnamefont{A.}~\bibnamefont{Lewis}},
  \bibinfo{author}{\bibfnamefont{A.}~\bibnamefont{Challinor}},
  \bibnamefont{and} \bibinfo{author}{\bibfnamefont{A.}~\bibnamefont{Lasenby}},
  \bibinfo{journal}{Astrophys. J.} \textbf{\bibinfo{volume}{538}},
  \bibinfo{pages}{473} (\bibinfo{year}{2000}), \eprint{astro-ph/9911177}.

\bibitem[{\citenamefont{Bassi and Ghirardi}(2003)}]{bassi2003}
\bibinfo{author}{\bibfnamefont{A.}~\bibnamefont{Bassi}} \bibnamefont{and}
  \bibinfo{author}{\bibfnamefont{G.~C.} \bibnamefont{Ghirardi}},
  \bibinfo{journal}{Phys.Rept.} \textbf{\bibinfo{volume}{379}},
  \bibinfo{pages}{257} (\bibinfo{year}{2003}), \eprint{quant-ph/0302164}.

\bibitem[{\citenamefont{Le\'{o}n and Sudarsky}(2015)}]{bispec}
\bibinfo{author}{\bibfnamefont{G.}~\bibnamefont{Le\'{o}n}} \bibnamefont{and}
  \bibinfo{author}{\bibfnamefont{D.}~\bibnamefont{Sudarsky}},
  \bibinfo{journal}{JCAP} \textbf{\bibinfo{volume}{1506}}, \bibinfo{pages}{020}
  (\bibinfo{year}{2015}), \eprint{1503.01417}.

\bibitem[{\citenamefont{Grishchuk and Sidorov}(1990)}]{grishchuk}
\bibinfo{author}{\bibfnamefont{L.}~\bibnamefont{Grishchuk}} \bibnamefont{and}
  \bibinfo{author}{\bibfnamefont{Y.}~\bibnamefont{Sidorov}},
  \bibinfo{journal}{Phys. Rev.} \textbf{\bibinfo{volume}{D 42}},
  \bibinfo{pages}{3413} (\bibinfo{year}{1990}).

\end{thebibliography}
\bibliographystyle{apsrev}

\end{document}